\documentclass[journal]{IEEEtran}
\usepackage{mathrsfs}
\usepackage{bbm}
\usepackage{amssymb}
\usepackage{bbding}
\usepackage{threeparttable}

\usepackage[mathcal]{euscript}

\usepackage{psfrag,calc,url,bm}

\usepackage{cite}

\usepackage{graphicx}

\usepackage{psfrag}

\usepackage{subfigure}

\usepackage{url}

\usepackage{stfloats}

\usepackage{amsmath}

\usepackage{float}

\usepackage{algorithmic}

\usepackage[ruled,linesnumbered,vlined]{algorithm2e}

\usepackage{color}

\usepackage{boxedminipage}

\usepackage{amsthm}

\usepackage{multirow}

\usepackage{setspace}

\usepackage{soul}
\usepackage{epstopdf}

\allowdisplaybreaks[3]

\usepackage{multirow}
\usepackage{amsmath}
\usepackage{booktabs}

\usepackage{makecell}

\begin{document}

\title{ICGNN: Graph Neural Network Enabled Scalable Beamforming  for MISO Interference Channels}
\author{Changpeng He, Yang Lu,~\IEEEmembership{Member,~IEEE}, Bo Ai,~\IEEEmembership{Fellow,~IEEE}, Octavia A. Dobre,~\IEEEmembership{Fellow,~IEEE}, \\Zhiguo  Ding,~\IEEEmembership{Fellow,~IEEE}, and Dusit Niyato,~\IEEEmembership{Fellow,~IEEE}
\thanks{Changpeng He and Yang Lu are with the School of Computer Science and Technology, Beijing Jiaotong University, Beijing 100044, China (e-mail: 24120332@bjtu.edu.cn, yanglu@bjtu.edu.cn).}
\thanks{Bo Ai is with the School of Electronics and Information Engineering, Beijing Jiaotong University, Beijing 100044, China (e-mail: boai@bjtu.edu.cn).}
\thanks{Octavia A. Dobre is with the Faculty of Engineering and Applied Science, Memorial University, St. John’s, NL A1C 5S7, Canada (E-mail: odobre@mun.ca).}
\thanks{Zhiguo Ding is with Department of Electrical Engineering and Computer Science, Khalifa University, Abu Dhabi 127788, UAE (e-mail: zhiguo.ding@ieee.org).}
\thanks{Dusit Niyato is with the College of Computing and Data Science, Nanyang Technological University, Singapore 639798 (e-mail: dniyato@ntu.edu.sg).}
}

\maketitle

%The hybrid maximum ratio transmission and zero-forcing scheme  is extended to reformulate the problem.

\begin{abstract}
This paper investigates the graph neural network (GNN)-enabled beamforming design for interference channels. We propose a model termed interference channel GNN (ICGNN) to solve a quality-of-service constrained energy efficiency  maximization problem. The ICGNN is two-stage, where the direction and power parts of beamforming vectors are learned separately but trained jointly via unsupervised learning. By formulating the dimensionality of features independent of the transceiver pairs, the ICGNN is scalable with  the  number of transceiver pairs. Besides, to improve the performance of the ICGNN, the hybrid maximum ratio transmission and zero-forcing scheme reduces the output ports, the feature enhancement module unifies the two types of links into one type, the subgraph representation  enhances the message passing efficiency, and the multi-head attention and residual connection facilitate the feature extracting. Furthermore, we present the over-the-air  distributed implementation of the ICGNN. Ablation studies validate the effectiveness of key components in the ICGNN. Numerical results also demonstrate the capability of ICGNN in achieving near-optimal performance with an average inference time less than $0.1$ ms. The scalability of ICGNN for unseen problem sizes is evaluated and enhanced by transfer learning with limited fine-tuning cost. The results of the centralized and distributed implementations of ICGNN are illustrated.
\end{abstract}


\section{INTRODUCTION}

\subsection{Background}

%multiple-input single-output (MISO)

With the fast- and ever-increasing traffic demand for data rates of mobile intelligent applications, the wireless communication systems have experienced tremendous growth and evolution with improved reliability and enhanced coverage \cite{6G}. Nevertheless, the scarcity of spectrum resource is still the key factor to limit the service capability of the wireless communication systems, which results in the non-orthogonal utilization of the  spectrum resource\cite{intro2}. Therefore, interference management becomes the central consideration in the development of future wireless communication systems. A classical interference channel system where multiple transmitter-receiver (Tx-Rx) pairs share the same spectral bandwidth have been regarded as a classical model to analyze and investigate the interference management \cite{intro3}. When employing the multi-antenna technology in interference channel systems, the beamforming design is capable of effectively managing interference by concentrating the radio frequency signal generated by one Tx towards its desired Rx while mitigating interference to other Rxs. However, finding the optimal beamforming design is usually shown to be an NP-hard problem. Although some heuristic algorithms or convex (CVX) optimization algorithms have been reported to derive the (near-)optimal beamforming design for various interference channel systems, the iteration frameworks inside these algorithms induce high computational complexity. Thus, it is difficult to realize real-time signal processing for time-varying wireless environment and dynamic network topology\cite{intro4}.

% \textcolor{blue}{Heuristic algorithms and convex (CVX) optimization algorithms have been used to derive (near-)optimal beamforming designs for various interference channel systems. However, these algorithms rely on iterative frameworks that result in high computational complexity. This complexity poses challenges for real-time signal processing in wireless environments with time-varying conditions and dynamic network topologies\cite{intro4}.}

% Traditional approaches to beamforming design often rely on optimization techniques such as successive convex approximation (SCA) \cite{SCA for beamforming} and block successive upper bound minimization (BSUM) \cite{BSUM for beamforming}. However, these methods typically require high computational complexity due to iterations, . Moreover, the beamforming design problems become extremely complicated for large-scale wireless networks \cite{high complexity iterative}.
%especially in dynamic environments with a large number of users.

Recently, some works have attempted to apply deep learning (DL) to address resource allocation problems for wireless networks \cite{intro5}, which shows superior performance than the traditional algorithms. In particular, neural networks can be trained to learn any intricate function, such as the mapping from channel state information  (CSI) to optimal power allocation scheme, to approximate and ``store" the solution capabilities of the traditional algorithms. For instance, fully connected multi-layer perceptron (MLP) and convolutional neural network (CNN) were respectively utilized to learn the non-linear mapping between inputs and outputs of beamforming design for next-generation full-duplex cellular systems \cite{MLP for beamforming design} and underwater acoustic communications\cite{CNN for beamforming design}. It is observed that both MLP and CNN are able to realize real-time and near-optimal inference. However, these models often struggle to capture inherent structure and relationships in wireless networks with graph-structured typologies \cite{Graph structure}. Graph neural network (GNN) has emerged as a powerful tool for processing data with complex relationships and dependencies \cite{Graph neural networks survey}. By representing wireless networks as graphs, where nodes correspond to users or base stations and edges represent channel links, GNN can inherently capture the topology features and interactions within the wireless networks. This graph-based learning approach has been shown to not only outperform other DL models in terms of beamforming learning but also scale with network typologies \cite{lugnn}, which shall be a promising signal processing strategy for future wireless systems. Additionally, the scalability of GNN plays a central role to realize distributed implementation for interference channel systems.

%Scale

% xxxx are trained to xxx （解决问题） for xxx （场景）.

 %making them attractive for real-time beamforming applications.\cite{Deep learning for wireless communications}.

% xxx工作提出一个xxx方法，解决什么问题，面向什么系统（顺序可以打乱），有什么发现。
%统一

% vs. single-antenna: real-value computaion
% vs. End-to-end learning: inference efficiency
% vs. archtecture design: attention
% vs. scale/distributed

%D2D: real-vale computaion -- beam: complex-value
%end-to-end learning: complex-value 但是task变复杂了。为了提升学习能力，attention..。但是，这些机制引入或者设计会增加计算量，推断时间与time-varying wireless enviroment不match。如何保障学习能力，又推断效率高成为一个关键问题。

%On the other hand, distributed. OTA 分布式的部署。但是，plug-and-play,不仅需要OTA，还需要模型具备对Tx-Rx pairs.现有GNN支持一定的扩展。例如，支持哪些类型的扩展。但是这些模型不能够直接扩展Tx-Rx pairs。

\subsection{Related Works}
Recently, GNN has made significant progress as a resource allocation enabler for wireless networks. For instance, in \cite{rw1}, the authors proposed a GNN-based meta-learning model aimed at solving the adaptive power control problem for dynamic device-to-device (D2D) communication systems. This model addressed the low adaptability and scalability issues of conventional DL methods by leveraging graph convolution and meta-learning approaches. In \cite{rw3}, the authors proposed a random edge graph neural network (REGNN)-based resource allocation strategy with the aim to maximize the weighted sum rate for a single-antenna interference channel system, where the property of permutation invariance of REGNN was retained to facilitate the training and implementation of REGNN in large-scale networks. In \cite{rw2}, the authors extended the REGNN to solve a set of system utility maximization problems for wireless ad-hoc or cellular networks. However, these models were designed based on real-value domains and cannot be directly applied to problems involving complex values, such as the multi-antenna beamforming design. To facilitate the GNN-based models to handle the complex-value problems, in \cite{cw1}, the authors proposed an unsupervised vanilla GNN-based model to solve the beamforming design for sum-rate maximization in multi-antenna D2D wireless networks. Instead of direct beamforming learning, the vanilla GNN-based model learned primal power and dual variables to reduce the learning complexity. In \cite{cw4}, the authors developed a message passing GNN (MPGNN) to solve the sum-rate maximisation problem for scalable radio resource management. In \cite{cw5}, the authors introduced a GNN-based model to maximize the system utilities by optimizing the beamforming vectors and  reflection phase shifts for an intelligent reflecting surface-aided system. However, the above models are all based on vanilla GNN. To enhance the feature extraction, in  \cite{dl}, a graph attention network (GAT)-based model was proposed to achieve energy-efficient beamforming design for multi-user MISO (MU-MISO) networks, where the attention mechanism was adopted to capture the hidden interactions among links. The GAT was shown to outperform vanilla GNN in terms of end-to-end beamforming learning. In \cite{dl3}, the authors devised a complex edge GAT (CEGAT) model to solve the max-min fair beamforming problem for MU-MISO networks, which took the edge features into account. From mentioned works, a trade-off is observed that complicated model architecture may enhance the expressive capability of GNN-based models while also increasing computational complexity. Therefore, jointly improving learning ability and inference efficiency becomes a central consideration for wireless networks as the channel conditions and network topology are time-varying.

On the other hand, researchers made efforts to apply GNN to wireless networks to achieve distributed deployment which requires the GNN-based resource allocation scalable with network typologies. For instance, in \cite{dl2}, the authors proposed a complex residual GAT (CRGAT) to solve the sum-rate maximization problem for MU-MISO networks, and the CRGAT was shown to scale with the numbers of users and power budgets. In \cite{sc2}, the authors presented a GNN-based model to solve the power allocation optimization problem for D2D cellular networks, which achieved scalability with network typologies and user quantities. In \cite{sc3}, the authors proposed a bipartite GNN (BGNN) to optimize the beamforming vectors for MU-MISO downlink networks. The BGNN partitioned the beamforming optimization process into multiple sub-operations dedicated to individual antenna vertices and user vertices to realize scalable beamforming calculation. In \cite{sc4}, the authors proposed a novel edge-GNN model to address the cooperative beamforming problem. The edge-GNN incorporated an edge-update process to enable the learning of cooperative beamforming on graph edges, which support scalability with the number of users and antennas. The above works reveal that the scalability  of GNN mainly depends on the parameter sharing. Moreover, the distributed deployment of GNN also requires information interaction and distributed computation. In \cite{dt1}, the authors presented an aggregation GNN (Agg-GNN) to realize decentralized resource allocation for a cooperative wireless system, where the Agg-GNN accepted the input of a sequence of local graph-aggregated state information obtained by each Tx from its multi-hop neighbors. In \cite{dt2}, the authors proposed a graph convolutional networks (GCN) based model to solve a distributed maximum weighted independent set (MWIS) problem for wireless multi-hop networks with orthogonal access, which was shown to generalize well across different types of graphs and utility distributions. In \cite{dt3}, the authors developed a distributed link specification model based on GCN with the aim to solve the distributed scheduling problem for dense wireless multi-hop networks. The model can significantly reduce scheduling overhead while retaining most of the network capacity by generating topology-aware and reusable node embedding parameters. In \cite{ota}, the authors proposed two models, i.e., air message passing neural network (Air-MPNN) and air message passing recurrent neural network (Air-MPRNN) to solve the distributed power allocation problem for wireless networks with dense Tx-Rx pairs. The two models leveraged the concept of over-the-air (OTA)  computation in message passing while Air-MPRNN reduced signaling overhead by utilizing graph embedding and local state from the previous frame to update the graph embedding in the current frame. Nevertheless, none of above models supports the ``plug-and-play" implementation\footnote{The ``plug-and-play" implementation indicates that the new Tx-Rx pair directly adopts an existing DL model and exchanges required signalling with exiting Tx-Rx pairs to realize distributed training  or  resource allocation.} of a new Tx-Rx pair to an existing wireless network.

\subsection{Contributions}
\emph{To the best of our knowledge, the DL model with the scalability with Tx-Rx pairs for MISO interference channels has not been reported in the existing works.} Note that such a scalability is very important to facilitate the ``plug-and-play" implementation as the number of Tx-Rx pairs changes in the system. To fill this gap, this paper proposes a GNN-based model for beamforming design in MISO interference channels, termed interference channel GNN (ICGNN). Particularly, we formulate and solve the quality-of-service (QoS) constrained EE maximization problem. The major contributions are summarized as follows:
\begin{itemize}
    \item We extend the hybrid maximum ratio transmission (MRT) and zero forcing (ZF) scheme proposed in \cite{hyb} to reformulate the considered problem with the aim to improve the learning performance. With the hybrid MRT and ZF scheme, the output ports of the desired mapping from CSI vectors to beamforming vectors are greatly reduced with limited or negligible performance loss.
    \item We develop the target mapping by the proposed ICGNN. Particularly, the feature enhancement unifies the two types of links (desired transmission and interference) into one type to facilitate the feature extraction. We implement the beamforming learning via two stages, i.e., the direction learning and the power learning, which are respectively realized by two GAT-based models. In each stage, the MISO interference channels are correspondingly represented by graphs and then, fed into models. We adopt the multi-head attention and residual connection to enhance the message passing to capture the complex relationships among links and mitigate the over-smoothing issue due to stacking deep layers. Finally, the learned power and direction features are utilized to recover the desired beamforming vectors. The effectiveness and scalability of the key modules in ICGNN are analyzed.
%Specifically, we employ two GAT-based learning modules to separately learn the direction and power of the beamforming vectors, and then recover the complete beamforming vectors based on a hybrid beamforming scheme. In the direction learning module, we introduce a novel graph representation to enhance the learning process, significantly improving the model's performance. This dual learning module approach, combined with hybrid beamforming scheme, enables ICGNN to effectively capture the complex relationships within the interference channel, resulting in more accurate and efficient beamforming vector generation.  Additionally, we incorporate residual structures to mitigate the over-smoothing problem. The integration of residual connections not only addresses the challenge of over-smoothing in deep GNN but also facilitates the propagation of important features across layers, leading to more robust and generalizable solutions for diverse channel conditions.
%    \item To better capture the complex characteristics of the MISO interference channels, we incorporate a feature enhancement module in the ICGNN model. This module processes the CSI to generate enriched and effective features that improve the performance of the subsequent GAT. The enhanced features help in better representation and learning, leading to more accurate beamforming vectors.

    \item The proposed ICGNN can be trained via unsupervised learning in both centralized and distributed manners. The power budget and QoS constraints are respectively guaranteed hardly by the activation function and softly by the penalty method. Particular, the OTA approach is adopted to realize the distributed implementation of the ICGNN following by the analysis of signaling overhead. Moreover, to reduce the signaling overhead, the \emph{one-value} OTA distributed learning is proposed.
\end{itemize}

We provide extensive simulation results to evaluate the proposed ICGNN. We conduct the ablation experiment to show the effectiveness of message passing, residual connection, graph representation, feature enhancement and hybrid MRT and ZF scheme to demonstrate their enhancement to the ICGNN. The optimality performance, scalability performance and feasibility rate of the ICGNN are validated. Especially, the ICGNN is shown to be scalable with the number of Tx-Rx pairs even though the problem size is unseen in the training set. We also showcase that the transfer learning is able to bridge performance gap with slight fine-tuning cost for the scenarios with large generalization error. Finally, we compare the centralized implementation and distributed implementation to reveal the trade-off between the learning performance and the signaling overhead.

%I propose a more comprehensive GNN-based approach to holistically address these challenges, called Interference Channel Graph Neural Network (ICGNN), for beamforming design in MISO interference channels.

The rest of this paper is organized as follows. Section II provides the system model and problem formulation. Section III presents the proposed ICGNN. Section IV describes the OTA implementation of the ICGNN. Section V presents  numerical results and performance analysis. Finally, Section VI concludes the paper.

\emph{Notation}: The following mathematical notations and symbols are used throughout this paper. $\bf a$ and $\bf A$ stand for a column vector and a matrix (including multiple-dimensional tensor), respectively. The sets of $n$-dimensional real column vector, $n$-by-$m$ real matrices and $n$-by-$m$-by-$k$ real tensors are denoted by ${\mathbb{R}^n}$, ${\mathbb{R}^{n \times m}}$ and ${\mathbb{R}^{n \times m \times k}}$, respectively. The sets of complex numbers, $n$-dimensional complex column vector and $n$-by-$m$ complex matrices are denoted by ${\mathbb{C}}$, ${\mathbb{C}^n}$ and ${\mathbb{C}^{n \times m}}$, respectively. For a complex number $a$, $\left| a \right|$ denotes its modulus and ${{\rm Re}(a)}$ denote its real part. For a vector $\bf a$, ${\left\| \bf a \right\|}$ is the Euclidean norm. For a matrix ${\bf A}$, ${\bf A}^H$ and $  \left \|{\bf A}\right\|$ denote its conjugate transpose and Frobenius norm, respectively. $[{\bf A}]_{i,j}$ and $[{\bf A}]_{i,:}$  are the $i$-th row and the $j$-th column element on the matrix $\bf A$ and the $i$-th row vector of the matrix $\bf A$, respectively. For a tensor ${\bf A}$, $[{\bf A}]_{i,:,:}$ denotes the matrix with index $i$ in the first dimension of the tensor.

\section{Problem Definition}

\subsection{System Model and Problem Formulation}

\begin{figure}[t]
\begin{center}
\centerline{\includegraphics[ width=.45\textwidth]{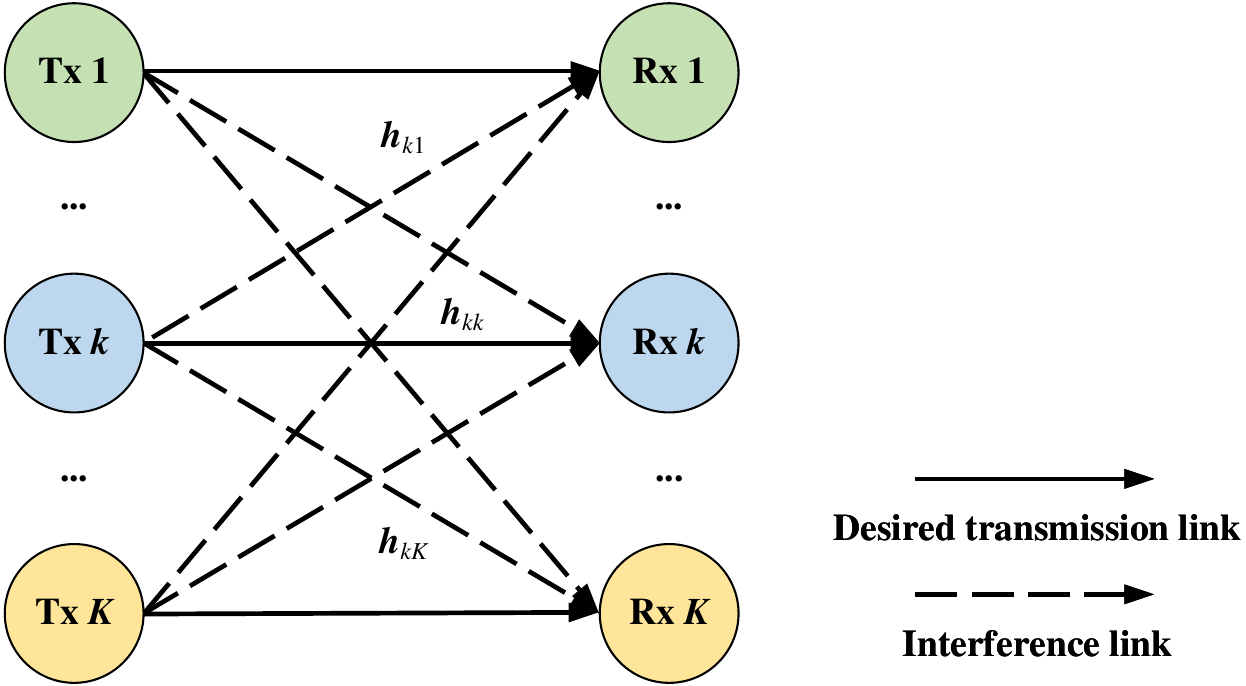}}
\caption{System model.}
\label{sys}
\end{center}
\end{figure}

Consider a MISO interference channel system\footnote{We use ``interference channels" in the context to represent the system where multiple Tx-Rx pairs operate over the same spectral bandwidth  \cite{Interference Channel}.} model as shown in Fig. \ref{sys}, where $K$ Tx-Rx pairs operate over a common spectrum. The Tx $k$ ($k\in\mathcal{K} \triangleq \{1,2, ..., K\}$) serves the Rx $k$ via their desired transmission link while introducing interference to other RXs via interference links meanwhile. The number of antennas of each Tx is $N_{\rm T}$ while each RX is single-antenna. %For clarity, we use $\mathcal{K} \triangleq \{1,2, ..., K\}$ to denote the index set of Tx-Rx pairs.

Denote the symbol and the corresponding beamforming vector for the Rx $k$ by $s_{k}\in\mathbb{C}$ and ${{{\bf{w}}_{k}}}\in{\mathbb{C}^{{N_{\rm{T}}}}}$, respectively. The received signal at the Rx $k$ is given by
\begin{flalign}
\label{AU_signal}
{{ y}_{k}}&=  {{\bf{h}}_{kk}^H{{\bf{w}}_{k}}{{s}_{k}}}+  {\sum\nolimits_{i=1,i \ne k}^K {{\bf h}_{ik}^H{{\bf w}_{i}}{{s}_{i}}} }  + {n_{k}},
\end{flalign}
where ${{{\bf{h}}_{ik}}}\in{\mathbb{C}^{{N_{\rm{T}}}}}$ denotes the CSI from the Tx $i$ to the Rx $k$ and $n_{k}\sim\mathcal{CN}( {0,{\sigma_{k} ^2}})$ denotes the additive white Gaussian noise (AWGN) at the Rx $k$. Without loss of generality, it is assumed that ${{\mathbb E}}\{ {{{| {s_{k} } |}^2}} \} = 1$ ($\forall k$). Then, the achievable information rate at the Rx $k$ is expressed as
\begin{flalign}\label{rate}
{R_k}\left( {\left\{ {{{\bf{w}}_i}} \right\}} \right) = {\log _2}\left( {1 + \frac{{{{\left| {{\bf h}_{kk}^H{{\bf w}_k}} \right|}^2}}}{{\sum\nolimits_{i = 1,i \ne k}^K {{{\left| {{\bf h}_{ik}^H{{\bf w}_i}} \right|}^2}} +\sigma _k^2}}} \right).
\end{flalign}

The QoS-constrained EE maximization problem is formulated as
\begin{subequations}\label{p0}
\begin{align}
&\mathop {\max }\limits_{\left\{{{{{\bf{w}}_{i}}}} \right\}} \frac{\sum\nolimits_{k = 1}^{{K}} {R_k}\left( {\left\{ {{{\bf{w}}_i}} \right\}} \right)}{\sum\nolimits_{k = 1}^{{K}} \|{\bf w}_k\|_2^2 + {P_{\rm c}}} \\
{\rm s.t.}~& {R_k}\left( {\left\{ {{{\bf{w}}_i}} \right\}} \right) \ge {R_{\rm req}} ,\label{cons:p0:A}\\
&{\left\| {{{\bf{w}}_{k}}} \right\|_2^2}  \le {P_{\rm max}},\label{cons:p0:B}\\
&{{\bf{w}}_{i}}\in{\mathbb{C}^{{N_{\rm{T}}}}}, {\forall i,k}\in{\cal K},
\end{align}
\end{subequations}
where ${P_{\rm c}}$ denotes the circuit power, ${R_{\rm req}}$ denotes the information rate requirement of each Rx, and ${P_{\rm max}}$ denotes the power budget of each Tx.

Problem \eqref{p0} is non-convex without close-form solutions. Generally, Problem \eqref{p0} can be solved by iterative CVX optimization algorithms. However, the CVX optimization algorithms may not be computational efficiency to realize real-time optimization. These algorithms can be regarded as a mapping from $\{{\bf h}_{kj}\}$ to ${\{ {\bf w}_k \}}$. Therefore, one can construct a neural network to approximate the mapping to facilitate the real-time inference.

\subsection{Problem Reformulation based on Hybrid MRT and ZF Learning}

In stead of directly solving Problem \eqref{p0}, we reformulate it based on the hybrid MRT and ZF learning in order to improve the computational efficiency.

The $k$-th  beamforming vector is composed of the power part denoted by $p_k$ and the direction part denoted by ${{\overline{\bf{w}}}_k}$, i.e.,
\begin{flalign}\label{mmse}
{{\bf{w}}_k} = {\sqrt {p_k}}{{\overline{\bf{w}}}_k},~{\left\| {{{\overline {\bf{w}} }_k}} \right\|^2} = 1.
\end{flalign}

The hybrid MRT and ZF scheme intends to set the direction part by the linear combination of the MRT direction and the ZF direction, which is given by
\begin{flalign}\label{hybrid}
{\overline {\bf w}}_k \left(\alpha_k\right) = \frac{\alpha_k \frac{{\bf u}_k}{\|{\bf u}_k\|} + \left(1-\alpha_k\right)\frac{{\bf h}_{kk}}{\|{\bf h}_{kk}\|}}{\left\|\alpha_k \frac{{\bf u}_k}{\|{\bf u}_k\|} + \left(1-\alpha_k\right)\frac{{\bf h}_{kk}}{\|{\bf h}_{kk}\|}\right\|},
\end{flalign}
where $\alpha_k \in [0,1]$ denotes the hybrid coefficient, and ${\bf u}_k$ is the $k$-th column of
\begin{flalign}\label{U_q}
{\bf{U}}_k = \left\{ \begin{array}{l}
{\bf{G}}_k^H{\left( {{\bf{G}}_k{\bf{G}}_k^H} \right)^{ - 1}},~{\rm if}~{N_{\rm{T}}} \ge K\\
{\left( {{\bf{G}}_k^H{\bf{G}}}_k \right)^{ - 1}}{\bf{G}}_k^H,~{\rm if}~{N_{\rm{T}}} < K
\end{array} \right.,
\end{flalign}
where ${\bf G}_k\triangleq[{\bf h}_{k1}^H,{\bf h}_{k2}^H,...,{\bf h}_{kK}^H]$.

Based on \eqref{mmse} and \eqref{hybrid}, ${R_k}( {\{ {{{\bf{w}}_i}} \}} )$ is expressed as
\begin{flalign}
{R_k} & \left( {\left\{p_i,\alpha_i \right\}} \right)= \nonumber\\
& {\log _2}\left( {1 + \frac{{{{\left| {{\bf{h}}_{kk}^H{{\sqrt{p_k}}\overline {\bf{w}}_k}\left( \alpha_k\right)} \right|}^2}}}{{\sum\nolimits_{i=1,i \ne k}^K {{{\left| {{\bf{h}}_{ik}^H{{\sqrt{p_i}}\overline{\bf{w}}_i}\left( \alpha_i \right)} \right|}^2}}  + {\sigma_k ^2}}}} \right).
\end{flalign}
Then, Problem \eqref{p0} is reformulated as
\begin{subequations}\label{p1}
\begin{align}
&\mathop {\max }\limits_{\left\{p_i,\alpha_i \right\}} \frac{\sum\nolimits_{k = 1}^{{K}} {R_k}\left( {\left\{p_i,\alpha_i \right\}} \right)}{\sum\nolimits_{k = 1}^{{K}} p_k + {P_{\rm c}}} \\
{\rm s.t.}~& {R_k}\left( {\left\{p_i,\alpha_i \right\}} \right) \ge {R_{\rm req}} ,\label{cons:p1:A}\\
&p_k  \le {P_{\rm max}},\label{cons:p1:B}\\
&\alpha_k \in [0,1],\label{cons:p1:C}\\
&{\forall i,k}\in{\cal K}.
\end{align}
\end{subequations}
It is seen that the required output dimensionality of the neural network is reduced from complex $N_{\rm T}\times K$ for Problem \eqref{p0} to real $2K$ for Problem \eqref{p1}. According to \cite{hyb} and \cite{Jorswieck}, the performance loss due to hybrid MRT and ZF scheme is limited or even negligible.

Then, the neural network can first learn the mapping from $\{{\bf h}_{kj}\}$ to the desired $\{p_k,\alpha_k \}$ and then, recover ${\{ {\bf w}_k \}}$ based on \eqref{mmse} and \eqref{hybrid}.

\newtheorem{rem}{Remark}
%\begin{rem}
%hybrid
%\end{rem}

%Denote $\{p_k^{({\rm out})}\}$ as the output power part of the neural networks. The power part is adjusted according to (\ref{cons:p0:B}) by the following activated function, \textcolor[rgb]{1,0,1}{i.e., ${p_k} = {\rm{Sigmod}}\left( {p_k^{\left( {{\rm{out}}} \right)}} \right)$, where $\rm{Sigmod}$ represents the sigmod activation function}

\section{GNN-Enabled Beamforming Design}

\begin{figure*}[t]
\centering
    \includegraphics[ width=1\textwidth]{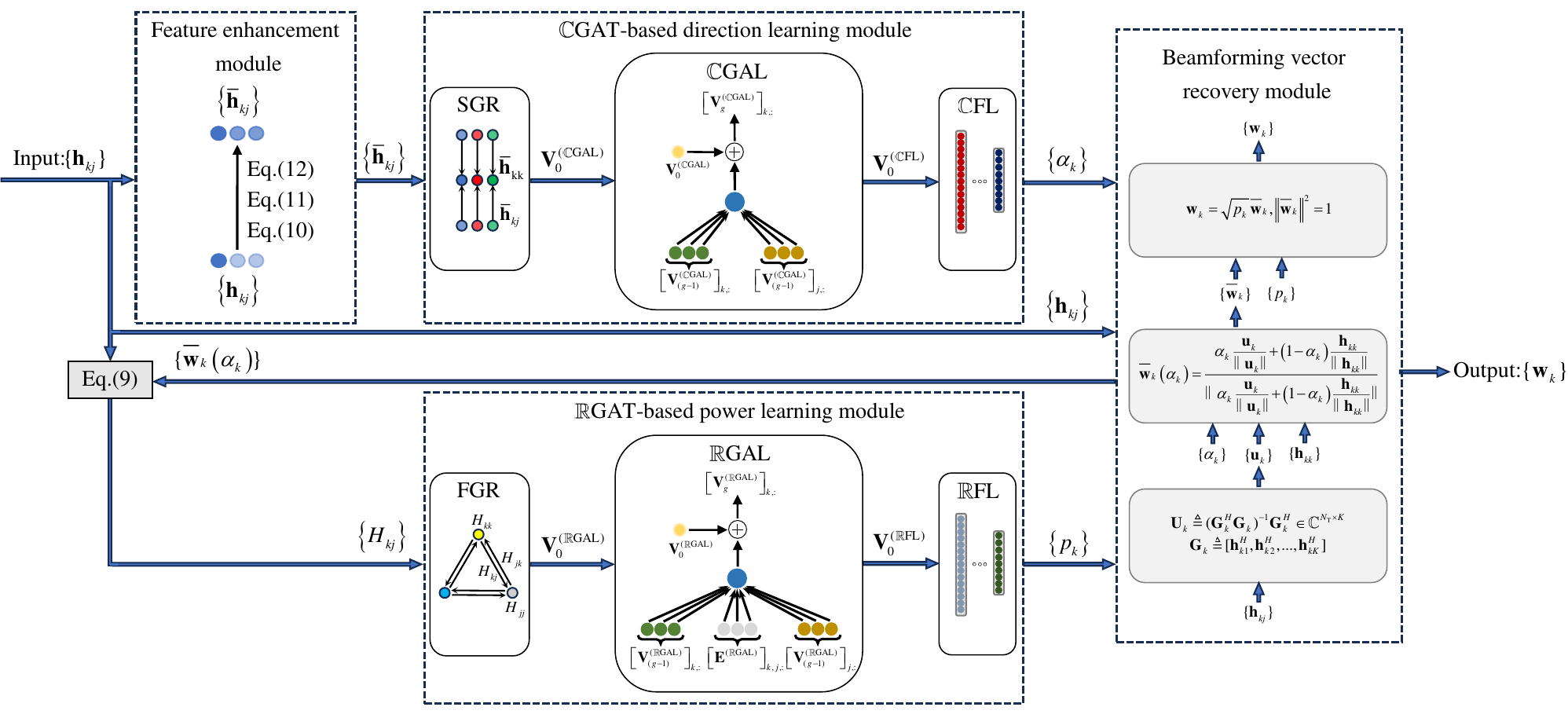}
    \caption{The architecture of the ICGNN: 1) feature enhancement module, 2)  ${\mathbb C}$GAT-based direction learning module, 3) ${\mathbb R}$GAT-based power learning module and 4) beamforming vector recovery module.}
    \label{Network Structure}
\end{figure*}

To solve Problem \eqref{p1}, we propose a GNN-enabled beamforming design for interference channels, termed ICGNN, as illustrated in Fig. \ref{Network Structure}, which includes four modules, i.e., feature enhancement module, ${\mathbb C}$GAT-based direction learning module, ${\mathbb R}$GAT-based power learning module, and beamforming vector recovery module.

In particular, the ICGNN is two-stage. During the first stage, all CSI is used to construct a tensor, i.e., $\{{\bf h}_{kj}\}$, which is enhanced by the feature enhancement module and then, input into the ${\mathbb C}$GAT-based direction learning module to generate $\{ {\alpha}_k \}$. During the second stage, we define $\{{{H}}_{{{kj}}}\}$ with the obtained $\{ {\alpha}_k \}$ by
\begin{flalign}\label{input2}
{{{H}}_{{{kj}}}} \triangleq {\left| {{\bf{h}}_{kj}^H{\overline {\bf w}}_k \left(\alpha_k\right)} \right|^2},~\forall k,j,
\end{flalign}
which is further input into the ${\mathbb R}$GAT-based power learning module to generate  $\{ {p}_k \}$. At last, the desired beamforming vectors are  derived via the beamforming vector recovery module with the obtained $\{ \alpha_k,{p}_k \}$.

Furthermore, to guarantee the output of the ICGNN to be feasible to Problem \eqref{p1}, the Sigmod function are adopted in the ${\mathbb C}$GAT-based direction learning module and the ${\mathbb R}$GAT-based power learning module following \eqref{cons:p1:B} and \eqref{cons:p1:C}. The QoS constraint \eqref{cons:p1:A} is applied into an unsupervised learning loss function with the objective function via the penalty method.

The basic information of four modules is summarized in Table \ref{table_s}, and the detailed processes and loss function are presented as follows.

\begin{table}[t]
  \begin{center}
    \caption{The input and output of the four modules of the ICGNN.}
     \label{table_s}
     \begin{tabular}{c || c|c}
     \hline
      Module & Input & Output  \\
      \hline
      \hline
      Feature enhancement & $\{{{\bf{h}}}_{kj}\}$ & $\{{\overline{\bf{h}}}_{kj}\}$ \\
      \hline
      ${\mathbb C}$GAT-based direction learning & $\{{\overline{\bf{h}}}_{kj}\}$ & $\{\alpha_k\}$ \\
      \hline
      ${\mathbb R}$GAT-based power learning& $\{{{H}}_{kj}\}$ & $\{p_k\}$  \\
      \hline
      Beamforming vector recovery& $\{{{\bf{h}}}_{kj},\alpha_k,p_k\}$ & $\{{\bf w}_k\}$  \\
      \hline
    \end{tabular}
 \end{center}
\end{table}

\subsection{Feature Enhancement Module}

In the considered problem, there are two types of links, i.e., the desired transmission link, i.e., $\{{\bf h}_{kk}\}$, and the interference link, i.e., $\{{\bf h}_{kj}\}_{k\ne j}$, which may result in heterogeneous graph-structured data in the graph representation. However, such the heterogeneity is hard to be captured during the message passing. Therefore, we propose a feature enhancement module to unify the two types of links into one new type, which makes the aggregation results of the subsequent ${\mathbb C}$GAT-based direction learning module more effective.

In particular, the feature enhancement module takes input of the CSI associated with each Tx, e.g., ${\bf G}_k$ for Tx $k$, and yields the corresponding enhanced features as
\begin{flalign}
{{{\bf{\overline h}}}_{kj}} = \left\{ {\begin{array}{*{20}{c}}
{{{\bf{h}}_{kk}}||{{\bf{h}}_{kk}},~k = j}\\
{{{\bf{h}}_{kj}}||{{{\bf{\widehat h}}}_{kj}},~k \ne j}
\end{array}} \right.,
\end{flalign}
where, $||$ denotes the concatenation operator and
\begin{flalign}
{{\widehat{\bf{h}}}_{kj}} = \frac{{\left\| {{{\bf{h}}_{kj}}} \right\|}}{{\left\| {{{{\bf{\mathord{\buildrel{\lower3pt\hbox{$\scriptscriptstyle\frown$}}
\over h} }}}_{kj}}} \right\|}}{{{\bf{\mathord{\buildrel{\lower3pt\hbox{$\scriptscriptstyle\frown$}}
\over h} }}}_{kj}},~k\ne j,
\end{flalign}
where
\begin{flalign}
{{{\bf{\mathord{\buildrel{\lower3pt\hbox{$\scriptscriptstyle\frown$}}
\over h} }}}_{kj}} = {{\bf{h}}_{kk}} - \frac{{{{{\bf{h}}_{kj}^H}{{\bf{h}}_{kk}}} }}{{ {{{\bf{h}}_{kk}^H}{{\bf{h}}_{kk}}} }}{{\bf{h}}_{kj}},~k\ne j.
\end{flalign}
Note that ${{\widehat{\bf{h}}}_{kj}}$ only adjusts the direction part of ${{{\bf{h}}}_{kj}}$ while $\|{{\widehat{\bf{h}}}_{kj}}\|=\|{{{\bf{h}}}_{kj}}\|$.

%The effectiveness of the feature enhancement module is analyzed in the following remark.

\begin{rem}\label{fem}
(Effectiveness of  feature enhancement) During the message passing, the two types of links are not distinguished explicitly. Direct aggregation of the features of two types of links can induce information loss and semantic ambiguity \cite{heterogeneit}. The main idea of the proposed feature enhancement is to convert the features of the interference links into features negatively correlated to interference level. For instance, ${{{\bf{\mathord{\buildrel{\lower3pt\hbox{$\scriptscriptstyle\frown$}}
\over h} }}}_{kj}}={\bf h}_{kk}$ if ${\bf h}_{kj}$ is perpendicular to ${\bf h}_{kk}$; ${{{\bf{\mathord{\buildrel{\lower3pt\hbox{$\scriptscriptstyle\frown$}}
\over h} }}}_{kj}}={\bf 0}_{N_{\rm T}}$ if ${\bf h}_{kj}$ is parallel to ${\bf h}_{kk}$. Moreover, the original features are also embedded into the enhanced features without missing the original information.
\end{rem}

\subsection{${\mathbb C}$GAT-based Direction Learning Module}

The ${\mathbb C}$GAT-based direction learning module is to map  $\{{{\overline{\bf{h}}}_{kj}}\}$ to $\{ {\alpha}_k \}$, which is composed of three components, i.e., subgraph representation (SGR), complex graph attention layer (${\mathbb C}$GAL), and complex fully-connected layer (${\mathbb C}$FL).

\subsubsection{Subgraph representation}

\begin{figure}[t]
\subfigure[Subgraph representation for direction learning.]{\label{graphical representation1} \includegraphics[ width=0.46\linewidth]{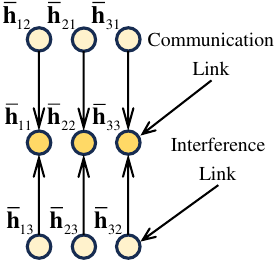}}
\quad
\subfigure[Fully-connected graph representation for power learning.]{\label{graphical representation2}  \includegraphics[ width=0.42\linewidth]{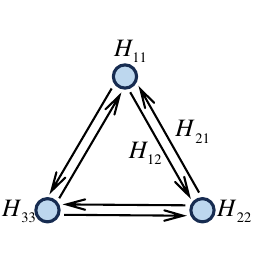}}
\caption{An example of graph representation with $K=3$ Tx-Rx pairs.}
\end{figure}

As shown in Fig. \ref{graphical representation1}, the considered MISO interference channel system  can be represented by a graph denoted by ${{\cal G}^{(1)}}$ which includes $K$ independent subgraphs associated with $K$ Txs. The $k$-th subgraph is denoted by ${\widetilde{\cal G}^{(1)}_k}=({\widetilde{\cal V}^{(1)}_k, {\widetilde{\cal E}}^{(1)}_k})$ where $\widetilde{\cal V}^{(1)}_k$ denotes the set of nodes and ${\widetilde{\cal E}}^{(1)}_k$ denotes the set of directed edges. In particular, each node in $\widetilde{\cal V}^{(1)}_k$ represents either the desired transmission link with the node feature of $\overline{\bf h}_{kk}$ or the interference link with the node feature of $\overline{\bf h}_{kj}$ $(j\ne k)$, and each edge in ${\widetilde{\cal E}}_k$ represents the existing relationship between two endpoints without edge features. Here, we use $\overline{{\bf h}}_{kj}$ to represent the features of $j$-th node in ${\widetilde{\cal G}^{(1)}_k}$ $(\forall k)$. Then, the adjacent matrix of ${\widetilde{\cal G}^{(1)}_k}$ is given by
\begin{flalign}
{\left[ {\bf{\Lambda }} \right]_{i,j}} = \left\{ {\begin{array}{*{20}{c}}
{1,{\text{ }}i \ne k{\text{ and }}j = k}\\
{0,{\text{\hspace{1.6em}otherwise\hspace{1.6em}}}}
\end{array}} \right..
\end{flalign}

\begin{rem} (Effectiveness of subgraph representation)
As shown in \eqref{hybrid}, the direction part of the $k$-th beamforming vector mainly depends on the local CSI information at the $k$-th Tx, i.e., ${\bf G}_k$. Accordingly, for the direction learning of ${\overline {\bf w}}_k (\alpha_k)$, only ${\bf G}_k$ is required. Therefore, we represent the MISO interference channels into  $K$ subgraphs, and the $k$-th subgraph is to learn ${\overline {\bf w}}_k (\alpha_k)$. Compared with traditional fully-connected graphs, such a SGR greatly reduces the exchanging and processing of irrelevant message during the message passing to enhance feature extracting efficiency.
%Compared with traditional fully-connected graphs, representing MISO interference channels as K subgraphs can enable GNN to only obtain channel information related to the k-th Tx. These channel information have the greatest impact on maximizing ${R_k}\left( {\left\{ {{{\bf{w}}_i}} \right\}} \right)$, while other information may also be effective but relatively small. So fully-connected graphs increase the difficulty of feature extraction during message passing due to receiving a lot of inefficient information, while subgraphs can focus on more effective information to reduce the difficulty of feature extraction.
\end{rem}

\subsubsection{Complex graph attention layer}

The ${\mathbb C}$GAL helps each node to extract features from the information from/to its neighboring nodes in the corresponding subgraph via message passing. To further enhance the learning capability, we adopt the attention mechanism.

We consider that $G_1$ ${\mathbb C}$GALs are employed. For the $g$-th ($g\in \{1,2,...,G_1\}$) ${\mathbb C}$GAL, denote ${\bf V}^{({{\mathbb C}\rm
 GAL})}_{g}\in {\mathbb C}^{K^2\times V_{1,g}}$ by the output node feature matrix, where $V_{1,g}$ represents the feature dimension of each node in the $g$-th ${\mathbb C}$GAL. That is, the input of the $g$-th ${\mathbb C}$GAL is ${\bf V}^{({{\mathbb C}\rm
 GAL})}_{(g-1)}$. Particularly, the input of the $1$-st ${\mathbb C}$GAL is ${\bf V}^{({{\mathbb C}\rm
 GAL})}_{0}$, which is given by
\begin{flalign}
{\left[ {{\bf{V}}_0^{({{\mathbb C}\rm
 GAL})}} \right]_{(k-1)\times K + j,:}} = {{\overline{\bf{h}}}_{kj}}.
\end{flalign}

Each ${\mathbb C}$GAL employs $D_1$ attention heads. The coefficient of the $d$-th ($d\in\{1,2,...,D_1\}$) attention head in the $g$-th
${\mathbb C}$GAL is denoted by ${\bf A}^{({{\mathbb C}\rm
 GAL})}_{g,d}\in {\mathbb R}^{K\times K}$, which is calculated by (\ref{attention coefficient0}), where ${\rm {\mathbb C}LeakyReLU}(\cdot)$ represents the complex LeakyReLU activation function, ${\bf W}^{({{\mathbb C}\rm
 GAL})}_{g,d}\in {\mathbb C}^{V_{1,(g-1)}\times V_{1,g}}$  and ${\bf a}^{({{\mathbb C}\rm
 GAL})}_{g,d}\in {\mathbb C}^{2V_{1,g}}$ denote the complex learning weights associated with the input node feature and attention in the $g$-th ${\mathbb C}$GAL, respectively, and ${{\cal N}_k}$ denotes the set of all neighboring nodes of node $k$.
\begin{figure*}
\begin{flalign}
\label{attention coefficient0}
&{[{{\bf{A}}^{({{\mathbb C}\rm
 GAL})}_{g,d}}]_{k,j}} = \frac{{\exp \left( {\left\| {{\mathop{\rm {\mathbb C}LeakyReLU}\nolimits} \left( {{{\bf{a}}^{({{\mathbb C}\rm
 GAL})}_{g,d}}^T \left( {\left[{\bf{V}}_{(g - 1)}^{({{\mathbb C}\rm
 GAL})}\right]_{k,:} {{\bf{W}}^{({{\mathbb C}\rm
 GAL})}_{g,d}}||\left[{\bf{V}}_{(g - 1)}^{({{\mathbb C}\rm
 GAL})}\right]_{j,:} {{\bf{W}}^{({{\mathbb C}\rm
 GAL})}_{g,d}}} \right)} \right)} \right\|} \right)}}{{\sum\nolimits_{i \in {{\cal N}_k}} {\exp } \left( {\left\| {{\mathop{\rm {\mathbb C}LeakyReLU}\nolimits} \left( {{{\bf{a}}^{({{\mathbb C}\rm
 GAL})}_{g,d}}^T\left( {\left[{\bf{V}}_{(g - 1)}^{({{\mathbb C}\rm
 GAL})}\right]_{k,:} {{\bf{W}}^{({{\mathbb C}\rm
 GAL})}_{g,d}}||\left[{\bf{V}}_{(g - 1)}^{({{\mathbb C}\rm
 GAL})}\right]_{i,:} {{\bf{W}}^{({{\mathbb C}\rm
 GAL})}_{g,d}}} \right)} \right)} \right\|} \right)}}
\end{flalign}
\hrule
\end{figure*}

Then, ${\bf V}^{({{\mathbb C}\rm GAL})}_{g}$ is calculated by \eqref{update1}, where ${\rm {\mathbb C}ReLU}(\cdot)$ represents the complex ReLU activation function, and ${\widehat{\bf W}}^{({{\mathbb C}\rm GAL})}_g\in {\mathbb C}^{V_{1,0}\times V_{1,g}}$ denotes the complex residual weight matrix. The residual weight is to realize the residual connection to mitigate the over-smoothing issue, which is realized by adding a linear mapping of the original node feature (i.e, ${{\bf{V}}_0^{({{\mathbb C}\rm GAL})}}$) to the extracted node feature.

 \begin{figure*}
\begin{flalign}\label{update1}
{\left[ {{\bf{V}}_g^{({{\mathbb C}\rm
 GAL})}} \right]_{k,:}}=& \mathop {||}\limits_{d = 1}^{D_1}  {\rm {\mathbb C}ReLU}\left(\sum\nolimits_{j \in \mathcal{N}_k} [{\bf A}^{({{\mathbb C}\rm
 GAL})}_{g,d}]_{k,j} {\left[ {{\bf{V}}_{g-1}^{({{\mathbb C}\rm
 GAL})}} \right]_{j,:}}{\bf W}^{({{\mathbb C}\rm
 GAL})}_{g,d} + {\left[ {{\bf{V}}_0^{({{\mathbb C}\rm
 GAL})}} \right]_{k,:}} {\widehat {\bf W}}^{({{\mathbb C}\rm
 GAL})}_{g,d} \right)
\end{flalign}
\hrule
\end{figure*}

\subsubsection{Complex fully-connected layer}

We consider that $F_1$ ${\mathbb C}$FLs are employed. For the $f$-th ($f\in\{1,2,...,F_1\}$) ${\mathbb C}$FL, denote ${\bf V}^{({{\mathbb C}\rm FL})}_{f}\in {\mathbb C}^{K\times H_{1,f}}$ by the output node features, where $H_{1,f}$ represents the feature dimension of each node in the $f$-th ${\mathbb C}$FL. Then, the input of the $1$-st ${\mathbb C}$FL is ${\bf V}^{({{\mathbb C}\rm FL})}_{0}$, which is given by
\begin{flalign}
 [{\bf V}^{({{\mathbb C}\rm FL})}_{0}]_{k,:}= [{\bf V}^{({{\mathbb C}\rm
 GAL})}_{G_1}]_{(k-1)\times K+k,:} ~.
\end{flalign}

For the  $f$-th ${\mathbb C}$FL with $f\in \{1,2,...,F_1-1\}$, the node feature matrix is updated by
\begin{flalign}
{{\bf V}^{({{\mathbb C}\rm FL})}_{f} = {\mathbb C}{\rm ReLU} \left({\bf V}^{({{\mathbb C}\rm FL})}_{\left(f-1\right)} {\bf W}^{({{\mathbb C}\rm FL})}_f + {\bf B}^{({{\mathbb C}\rm FL})}_f\right)},
\end{flalign}
where ${\bf W}^{({{\mathbb C}\rm FL})}_f \in {\mathbb C}^{H_{1,(f-1)}\times H_{1,f}}$ and ${\bf B}^{({{\mathbb C}\rm FL})}_f\in {\mathbb C}^{K^2\times H_{1,f}}$ denote the complex weight matrix and the complex bias, respectively. Note that $K^2$ row vectors in ${\bf B}^{({{\mathbb C}\rm FL})}_f$ are identical to guarantee scalability with the number of Tx-Rx pairs of the model. Furthermore, a complex BatchNorm (BN) \cite{BatchNorm} layer is added after each ${\mathbb C}$FL in order to prevent overfitting and enhance the convergence behavior.

For the $F_1$-th ${\mathbb C}$FL, the update process is given by
\begin{flalign}
{\bf v}^{({{\mathbb C}\rm FL})}_{F_1} =  {\bf V}^{({{\mathbb C}\rm FL})}_{\left(F_1-1\right)} {\bf w}^{({{\mathbb C}\rm FL})}_{F_1} + {\bf b}^{({{\mathbb C}\rm FL})}_{F_1},
\end{flalign}
where ${\bf w}^{({{\mathbb C}\rm FL})}_{F_1} \in {\mathbb C}^{H_{1,(F_1-1)}}$ and ${\bf b}^{({{\mathbb C}\rm FL})}_{F_1}\in {\mathbb C}^{K}$. Then, we adopt the Sigmod activation function to the real part of ${\bf v}^{({{\mathbb C}\rm FL})}_{F_1}$ to obtain the output of  the ${\mathbb C}$GAT-based direction learning module, i.e.,
\begin{flalign}\label{af2}
\left[\alpha_1,\alpha_2,...,\alpha_{K}\right]^T = \delta\left({\rm Re}\left( {\bf v}^{({{\mathbb C}\rm FL})}_{F_1} \right)\right).
\end{flalign}
Note that it holds that $\alpha_k \in (0,1)$ $(\forall k)$ via \eqref{af2}.

With the obtained $\{\alpha_k\}$, the MISO interference channel system can be equivalently transformed into a ``virtual" SISO interference channel system where the channel vectors, i.e., $\{{\bf h}_{kj}\}$ are represented by channel gains, i.e., $\{{H}_{kj}\}$ (cf. \eqref{input2}).

\subsection{${\mathbb R}$GAT-based Power Learning Module}

The ${\mathbb R}$GAT-based direction learning module is to map  $\{{H}_{kj}\}$ to $\{ p_k \}$, which is composed of three components, i.e., fully-connected graph representation (FGR), real graph attention layer (${\mathbb R}$GAL) and real fully-connected layer (${\mathbb R}$FL).

\subsubsection{Fully-connected graph representation} As shown in Fig. \ref{graphical representation2}, the ``virtual" SISO interference channel system is represented by a fully-connected graph denoted by ${{\cal G}_2}=({{\cal V}_2, {\cal E}_2})$, where ${\cal V}_2$ denotes the set of nodes and ${\cal E}_2$ denotes the set of directed edges. There are $K$ nodes which represent $K$ desired transmission links and $K(K-1)$ edges which represent $K(K-1)$ interference links. The feature of the $k$-th node is denoted by $H_{kk}$ while the feature of the $<k,j>$-th ($k\ne j$) edge is denoted by $H_{kj}$.

\subsubsection{Real graph attention layer}

The ${\mathbb R}$GAL is used for feature extraction over ${\cal G}_2$ to update its node features. Different from the ${\mathbb C}$GAL where the message passing is only based on the node features, the ${\mathbb R}$GAL adopts an edge-assisted message passing.

Assume that $G_2$ ${\mathbb R}$GALs are employed. For the $g$-th ($g\in \{1,2,...,G_2\}$) ${\mathbb R}$GAL, denote ${\bf V}^{({{\mathbb R}\rm
 GAL})}_{g}\in {\mathbb R}^{K\times V_{2,g}}$ and ${\bf E}^{({{\mathbb R}\rm
 GAL})}\in {\mathbb R}^{K\times K \times N_{\rm T}}$by the output node feature matrix and edge feature matrix, respectively, where $V_{2,g}$ represents the feature dimension of each node in the $g$-th ${\mathbb R}$GAL. That is, the input of the $g$-th ${\mathbb R}$GAL is $\{{\bf V}^{({{\mathbb R}\rm
 GAL})}_{(g-1)},{\bf E}^{({{\mathbb R}\rm
 GAL})}\}$. Particularly, the input of the $1$-st ${\mathbb R}$GAL is given by
\begin{flalign}
{\left[ {{\bf{V}}_0^{({{\mathbb R}\rm
 GAL})}} \right]_{k,:}} = {{H}}_{kk},
\end{flalign}
\begin{flalign}
{\left[ {{\bf{E}}^{({{\mathbb R}\rm
 GAL})}} \right]_{k,j,:}} = {{H}}_{kj}.
\end{flalign}

Each ${\mathbb R}$GAL employs $D_2$ attention heads. The coefficient of the $d$-th ($d\in\{1,2,...,D_2\}$)  attention head in the $g$-th
${\mathbb R}$GAL is denoted by ${\bf A}^{({{\mathbb R}\rm
 GAL})}_{g,d}\in {\mathbb R}^{K\times K}$, which is calculated by (\ref{attention coefficient}), where ${\rm {\mathbb R}LeakyReLU}(\cdot)$ represents the real LeakyReLU activation function, and ${\bf W}^{({{\mathbb R}\rm
 GAL})}_{g,d}\in {\mathbb R}^{V_{2,(g-1)}\times V_{2,g}}$, ${\overline {\bf W}}^{({{\mathbb R}\rm
 GAL})}_{g,d}\in {\mathbb R}^{N_{\rm T}\times V_{2,g}}$ and ${\bf a}^{({{\mathbb R}\rm
 GAL})}_{g,d}\in {\mathbb R}^{3V_{2,g}}$ denote the real learning weights associated with the input node feature, edge feature and attention mechanism of the $g$-th ${\mathbb R}$GAL, respectively.
\begin{figure*}
\begin{flalign}
\label{attention coefficient}
&{[{{\bf{A}}^{({{\mathbb R}\rm
 GAL})}_{g,d}}]_{k,j}} = \\
 & \frac{{\exp \left( { {{\mathop{\rm {\mathbb R}LeakyReLU}\nolimits} \left( {{{\bf{a}}^{({{\mathbb R}\rm
 GAL})}_{g,d}}^T \left( {\left[{\bf{V}}_{(g - 1)}^{({{\mathbb R}\rm
 GAL})}\right]_{k,:} {{\bf{W}}^{({{\mathbb R}\rm
 GAL})}_{g,d}}||\left[{\bf{V}}_{(g - 1)}^{({{\mathbb R}\rm
 GAL})}\right]_{j,:} {{\bf{W}}^{({{\mathbb R}\rm
 GAL})}_{g,d}}}||\left[{\bf E}^{({{\mathbb R}\rm
 GAL})}\right]_{k,j,:} {\overline {\bf W}}^{({{\mathbb R}\rm
 GAL})}_{g,d} \right)} \right)} } \right)}}{{\sum\nolimits_{i \in {{\cal N}_k}} {\exp } \left( { {{\mathop{\rm {\mathbb R}LeakyReLU}\nolimits} \left( {{{\bf{a}}^{({{\mathbb R}\rm
 GAL})}_{g,d}}^T\left( {\left[{\bf{V}}_{(g - 1)}^{({{\mathbb R}\rm
 GAL})}\right]_{k,:} {{\bf{W}}^{({{\mathbb R}\rm
 GAL})}_{g,d}}||\left[{\bf{V}}_{(g - 1)}^{({{\mathbb R}\rm
 GAL})}\right]_{i,:} {{\bf{W}}^{({{\mathbb R}\rm
 GAL})}_{g,d}}}||\left[{\bf E}^{({{\mathbb R}\rm
 GAL})}\right]_{k,i,:} {\overline {\bf W}}^{({{\mathbb R}\rm
 GAL})}_{g,d} \right)} \right)} } \right)}}\nonumber
\end{flalign}
\hrule
\end{figure*}

\begin{figure*}
\begin{flalign}\label{update2}
&{\left[ {{\bf{V}}_g^{({{\mathbb R}\rm
 GAL})}} \right]_{k,:}}= \mathop {||}\limits_{d = 1}^{D_2}  {\rm {\mathbb R}ReLU}\left(\sum\nolimits_{j \in \mathcal{N}_k} [{\bf A}^{({{\mathbb R}\rm
 GAL})}_{g,d}]_{k,j} {\left[ {{\bf{V}}_{(g-1)}^{({{\mathbb R}\rm
 GAL})}} \right]_{j,:}} {\bf W}^{({{\mathbb R}\rm
 GAL})}_{g,d} +  {\left[ {{\bf{V}}_0^{({{\mathbb R}\rm
 GAL})}} \right]_{k,:}} {\widehat {\bf W}}^{({{\mathbb R}\rm
 GAL})}_{g}\right)
\end{flalign}
\hrule
\end{figure*}

Then, ${\bf V}^{({{\mathbb R}\rm
 GAL})}_{g}$ is calculated by \eqref{update2}, where  ${\rm {\mathbb R}ReLU}(\cdot)$ represents the real ReLU activation function, and ${\widehat{\bf W}}^{({{\mathbb R}\rm
 GAL})}_g\in {\mathbb R}^{V_{2,0}\times V_{2,g}}$ denotes the residual weight matrix.

\subsubsection{Real fully-connected layer}

Assume that $F_2$ ${\mathbb R}$FLs are employed. For the $f$-th ($f\in\{1,2,...,F_2\}$) ${\mathbb R}$FL, denote ${\bf V}^{({{\mathbb R}\rm FL})}_{f}\in {\mathbb R}^{K\times H_{2,f}}$ by the output node features, where $H_{2,f}$ represents the feature dimension of each node in the $f$-th ${\mathbb R}$FL. The input of the $1$-st ${\mathbb R}$FL is ${\bf V}^{({{\mathbb R}\rm FL})}_{0}$, which is given by
\begin{flalign}
 {\bf V}^{({{\mathbb R}\rm FL})}_{0}= {\bf V}^{({{\mathbb R}\rm
 GAL})}_{G_2}.
\end{flalign}

For the  $f$-th ${\mathbb R}$FL with $f\in \{1,2,...,F_2-1\}$, the node feature matrix is updated by
\begin{flalign}
{{\bf V}^{({{\mathbb R}\rm FL})}_{f} = {\mathbb R}{\rm ReLU} \left({\bf V}^{({{\mathbb R}\rm FL})}_{\left(f-1\right)} {\bf W}^{({{\mathbb R}\rm FL})}_f + {\bf B}^{({{\mathbb R}\rm FL})}_f\right)},
\end{flalign}
where ${\bf W}^{({{\mathbb R}\rm FL})}_f \in {\mathbb R}^{H_{2,(f-1)}\times H_{2,f}}$ and ${\bf B}^{({{\mathbb R}\rm FL})}_f\in {\mathbb R}^{K\times H_{2,f}}$ denote the real weight matrix and the real bias, respectively. Note that $K$ row vectors in ${\bf B}^{({{\mathbb R}\rm FL})}_f$ are also set to be identical. Besides, a real BN layer is added after each ${\mathbb R}$FL.

For the $F_2$-th ${\mathbb R}$FL, the update process is given by
\begin{flalign}
{\bf v}^{({{\mathbb R}\rm FL})}_{F_2} =  {\bf V}^{({{\mathbb R}\rm FL})}_{\left(F_2-1\right)} {\bf w}^{({{\mathbb R}\rm FL})}_{F_2} + {\bf b}^{({{\mathbb R}\rm FL})}_{F_2},
\end{flalign}
where ${\bf w}^{({{\mathbb R}\rm FL})}_{F_2} \in {\mathbb R}^{H_{2,(F_2-1)}}$ and ${\bf b}^{({{\mathbb R}\rm FL})}_{F_2}\in {\mathbb R}^{K}$. Then, we adopt the Sigmod activation function to the ${\bf v}^{({{\mathbb R}\rm FL})}_{F_2}$ to obtain the output of  the ${\mathbb R}$GAT-based direction learning module, i.e.,
\begin{flalign}
\left[p_1,p_2,...,p_{K}\right]^T = \delta\left({\bf v}^{({{\mathbb R}\rm FL})}_{F_2} \right).
\end{flalign}
Note that it holds that $p_k \in (0,1)$ $(\forall k)$.

\subsection{Beamforming Vector Recovery Module}

The  beamforming vector recovery module is added to recover the beamforming vectors with the obtained $\{\alpha_k\}$ and $\{p_k\}$, as the goal of solving Problem \eqref{p0} is to find the beamforming vectors $\{{\bf w}_k\}$. Particularly, we first calculate $\{{\bf u}_k\}$ with $\{{\bf h}_k\}$ by (\ref{U_q}) and then, calculate $\{{\overline {\bf w}}_k (\alpha_k)\}$ with $\{{\bf u}_k\}$, $\{{\bf h}_{kk}\}$ and $\{\alpha_k\}$ by (\ref{hybrid}). As a result, $\{{\bf w}_k\}$ are recovered with $\{{\overline {\bf w}}_k(\alpha_k)\}$ and $\{p_k\}$ by (\ref{mmse}).

\subsection{Loss function and unsupervised learning}

Denote $\{{\bf w}^{(n)}| {\bm \theta}\}$ by the output of the ICGNN with the input of $\{{{\bf{h}}}^{(n)}\}$ and the learnable parameters of the ICGNN
\begin{flalign}
{\bm \theta}\triangleq&\left[{\bf a}^{({{\mathbb C}\rm
 GAL})}_{g,d}, {\bf W}^{({{\mathbb C}\rm
 GAL})}_{g,d}, {\widehat {\bf W}}^{({{\mathbb C}\rm
 GAL})}_{g,d}, {\bf W}^{({{\mathbb C}\rm
 FL})}_{f}, {\bf w}^{({{\mathbb C}\rm
 FL})}_{F}, \right. \nonumber  \\
 &{\bf B}^{({{\mathbb C}\rm
 FL})}_{f}, {\bf b}^{({{\mathbb C}\rm
 FL})}_{F}, {\bf a}^{({{\mathbb R}\rm
 GAL})}_{g,d}, {\bf W}^{({{\mathbb R}\rm
 GAL})}_{g,d}, {\bf {\overline W}}^{({{\mathbb R}\rm
 GAL})}_{g,d}, \nonumber \\
 &\left.{\widehat {\bf W}}^{({{\mathbb R}\rm
 GAL})}_{g,d}, {\bf W}^{({{\mathbb R}\rm
 FL})}_{f},  {\bf w}^{({{\mathbb R}\rm
 FL})}_{F}, {\bf B}^{({{\mathbb R}\rm
 FL})}_{f}, {\bf b}^{({{\mathbb R}\rm
 FL})}_{F}\right],
 \end{flalign}
 where $n$ represents the index of the $n$-th sample in the training set. Then, ${R_k}(\{{\bf w}^{(n)}_k |{\bm \theta} \})$ is obtained via (\ref{rate}) with $\{{\bf w}^{(n)}_k|{\bm \theta}\}$.

 % Then, ${{{\bf w}^{(n)}_k}}| {\bm \theta}$ can be obtained by calculating the equation (\ref{mmse}) and (\ref{hybrid}) with  ${{\alpha}^{(n)}_k}| {\bm \theta}$ and ${{p}^{(n)}_k}| {\bm \theta}$, and ${R_k}(\{{\bf w}^{(n)}_k |{\bm \theta} \})$ is obtained via (\ref{rate}) with ${\bf w}^{(n)}_k|{\bm \theta}$.

With a batch of $N$ samples, i.e., $\{{{\bf{h}}}^{(n)}\}_{n=1}^{N}$, the loss function used to update  ${\bm \theta}$ is given by (\ref{Loss_Function}), where a penalty term is added to improve the constraint satisfaction rate.
\begin{figure*}
\begin{flalign}\label{Loss_Function}
{{\cal L}_N}\left(\bm \theta  \right) = \frac{1}{N}\sum\nolimits_{n = 1}^N {\left( {\frac{{\sum\nolimits_{k = 1}^K {{{\left\| {\left\{ {{\bf{w}}_k^{(n)}|\bm \theta } \right\}} \right\|}^2}}  + {P_{\rm{C}}}}}{{\sum\nolimits_{k = 1}^K {{R_k}} \left( {\left\{ {{\bf{w}}_k^{(n)}|\bm \theta } \right\}} \right)}} + \sum\nolimits_{k = 1}^K {{\rm {\mathbb R}ReLU}\left( {{R_{{\rm{req}}}} - {R_k}\left( {\left\{ {{\bf{w}}_k^{(n)}|\bm \theta } \right\}} \right)} \right)} } \right)}
\end{flalign}
\hrule
\end{figure*}

\begin{algorithm}[t]
\caption{Training ICGNN.}
 {\bf Training dataset:}  $\{{\bf V}^{(m)} \}_{m=1}^{M}$\;
 {\bf Initialize}  $\bm\theta$\;
 \For{{\rm epoch} $e \in [0,1,\dots,E] $}{
 \For{{\rm minibatch} $b \in [0,1,\dots,B] $}{Sample the $b$-th minibatch $\{{\bf V}^{(n)} \}_{n=1}^{N}$\;
 Obtain hybrid coefficient $\{ {\alpha}_k \}$ and power of beamforming vector $\{ {p}_k \}$ via the ICGNN with $\bm \theta$\;
 Calculate the beamforming vectors $\{{{{\bf{w}}^{(n)}_k}}| {\bm \theta}\}_{n=1}^N$\ via (\ref{mmse}) and (\ref{hybrid})\;
 Calculate the loss function ${\cal L}_N(\bm \theta)$ via (\ref{Loss_Function})\;
 Update ${\bm \theta}$ via optimizer, e.g., Adam \cite{Adam}\;
 }
}
{\bf Return}  $\bm\theta$.
\label{UL}
\end{algorithm}

The overall training processes are summarized in Algorithm 1, where $M$, $E$ and $B$  denote the numbers of samples, epochs and batches, respectively.

\begin{rem}\label{sca} (Scale with the number of Tx-Rx pairs)
It is observed that only two learnable parameters in $\bm \theta$, i.e., ${\bf B}^{({{\mathbb C}\rm FL})}_{f}$ and ${\bf B}^{({{\mathbb R}\rm FL})}_{f}$, are dependent on the number of Tx-Rx pairs $K$. Nevertheless, as mentioned above, the $K^2$ row vectors of  ${\bf B}^{({{\mathbb C}\rm FL})}_{f}$ and the $K$ row vectors of ${\bf B}^{({{\mathbb R}\rm FL})}_{f}$ are identical, for any value of $K$. That is, all computations in the ICGNN are independent of $K$, which indicates that the ICGNN is scale with the number of Tx-Rx pairs.
%model demonstrates a remarkable ability to generalize across different numbers of Tx-Rx pairs. This scalability stems from the model's architecture, where only the dimension of the parameters ${\bf B}^{({{\mathbb C}\rm GAL})}_{f}$ and ${\bf B}^{({{\mathbb R}\rm GAL})}_{f}$, depends on the number of Tx-Rx pairs $K$. However, since the $K$ row vectors of ${\bf B}^{({{\mathbb C}\rm GAL})}_{f}$ and ${\bf B}^{({{\mathbb R}\rm GAL})}_{f}$ are identical, ${\bf B}^{({{\mathbb C}\rm GAL})}_{f}$ and ${\bf B}^{({{\mathbb R}\rm GAL})}_{f}$ is effectively independent of $K$. This design allows the model to adapt to scenarios with varying numbers of Tx-Rx pairs, even those unseen during the training process.
\end{rem}

\begin{rem}\label{transfer} (Transfer learning)
Note that although the ICGNN is scalable with the number of Tx-Rx pairs, the expressive performance may be degraded when the unseen problem size differs greatly from the training samples. In this case, the transfer learning \cite{dl2} can be adopted to fine-tune the ICGNN at little training cost to enhance its expressive capability in the new application scenarios.
\end{rem}

\section{Over-the-air Implementation}

The proposed ICGNN can be implemented in a centralized manner or a distributed manner.
\begin{itemize}
    \item Centralized implementation: All Txs feedback the collected CSI to a computing center and share the parameters of ICGNN during the training phase.
    \item Distributed implementation: Each Tx collects only its associated CSI, e.g., the Tx $k$ with $\{{\bf h}_{kj}\}_j$. The Txs exchange messages among each other to facilitate the training its individual ICGNN locally.
\end{itemize}

\begin{figure*}[t]
\centering
    \includegraphics[ width=1\textwidth]{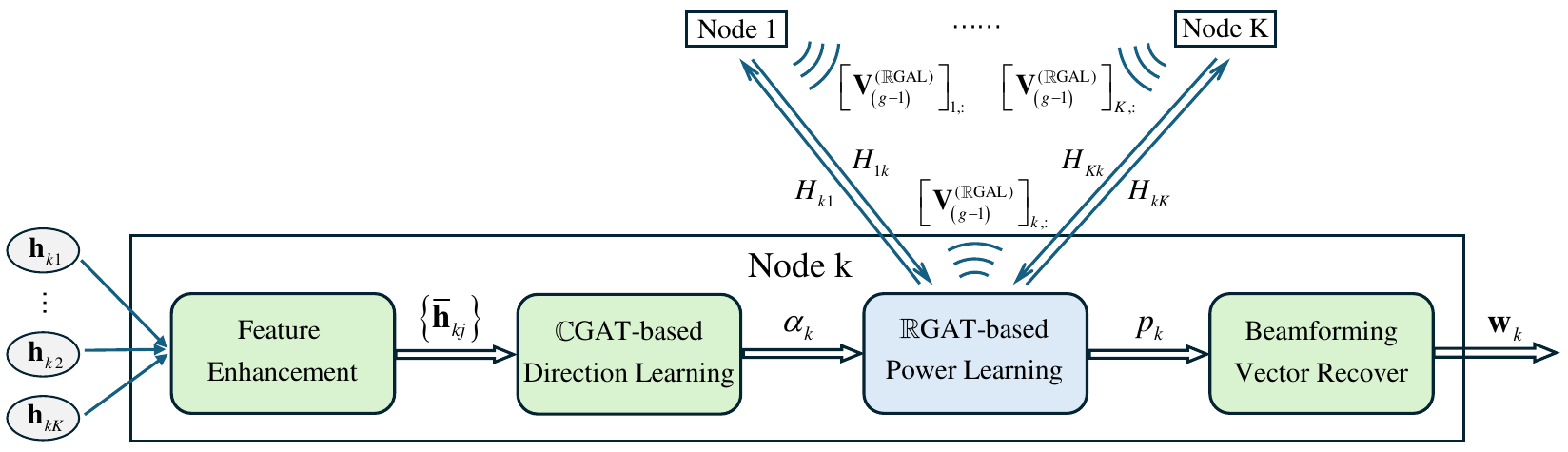}
    \caption{Illustration of the ICGNN via  OTA distributed implementation. The node (TX-Rx pair) $k$ collects its local CSI, i.e., $\{{\bf h}_{kj}\}_j$, and executes the first two modules of the ICGNN. During the power learning, it exchanges OTA messages with other nodes. Finally, it obtains the desired beamforming vector ${\bf  w}_k$.}
    \label{over-the-air implementation}
\end{figure*}

The distributed implementation can be realized via the OTA computing. Particularly, one module in the ICGNN requires messages from other nodes, i.e., the power learning module. Therefore, we leverage the OTA computation to facilitate the distributed implementation of the module which is illustrated in Fig. \ref{over-the-air implementation}.

After feature enhancement and direction learning locally, the $k$-th node obtains $\{{{{H}}_{{{kj}}}} \}_j$. Define ${\bf a}^{\rm OTA}_{k,(g-1)}$ as the message in the $g$-th round OTA message passing of the $k$-th node, i.e.,
\begin{flalign}
{\bf a}^{\rm OTA}_{k,(g-1)}=\left\{ \begin{array}{l}
[{{H}}_{k1},...,{{H}}_{kk},...,{{H}}_{kK}]^T\in {\mathbb R}^{K},~g=1,\\
{\left[ {{\bf{V}}_{\left(g-1\right)}^{({{\mathbb R}\rm
 GAL})}} \right]_{k,:}} \in {\mathbb R}^{K\times V_{2,\left(g-1\right)}},~g>1.
\end{array} \right.
\end{flalign}
In particular, in the $1$-st round message passing, the $k$-th node requires to broadcast its $H_{kk}$ to other $(K-1)$ nodes and sent $H_{kj}$ ($j\ne k$) to the $j$-th node. In the $g$-th ($g>1$) round message passing, the $k$-th node requires to broadcast its ${[ {{\bf{V}}_{(g-1)}^{({{\mathbb R}\rm
 GAL})}} ]_{k,:}}$ to other $(K-1)$ nodes, as the edge features are not updated in the ${\mathbb R}$GAT-based power learning module.

 Such message passing can be implemented via wired or wireless channel such as \cite{ota}. The difference lies in the signal processing to recover the OTA message. Nevertheless, the signaling overhead due to exchanging the OTA message is similar, which  consists of two parts, i.e., node features in all $G_2$ rounds (i.e., the number of ${\mathbb R}$GALs in the direction learning module) and edge features in the $1$-st round. Then, the total signaling overhead for the OTA implementation is given by
\begin{flalign}
\delta^{\left( {{\rm{OTA}}} \right)}_{\rm Total} =  K\sum\nolimits_{g=1}^{{G_2}} {\delta _g^{\left( {{\rm{OTA}}} \right)}} + K\left(K-1\right),
\end{flalign}
where ${\delta _g^{\left( {{\rm{OTA}}} \right)}}$ is the number of symbols of the OTA message in the $g$-th round of message passing which is related to $V_{2,\left(g-1\right)}$. Note that to reduce the signaling overhead, one can set $V_{2,\left(g-1\right)}=1$ $(\forall g\in\{1,...,G_2\})$, which results in \emph{one-value OTA} message passing. Intuitively, the one-value OTA distributed learning may reduce the signaling overhead at the expense of learning performance degradation.

With the OTA implementation, one Tx-Rx pair can join a interference channel system via a ``plug-and-play" protocol with the following steps.
\begin{enumerate}
\item The Tx-Rx pair registers itself with a server of the system to implement the ICGNN as well as reporting its presence to and synchronizing with other TX-Rx pairs.

\item The Tx processes channel estimation to obtain its local CSI and executes the ICGNN. Particularly, it exchanges OTA messages with other Tx-Rx pairs during the power learning.

\item The Tx obtains the desired beamforming vector for its Rx with the ICGNN.

\end{enumerate}
It is observed that the scalability with  the number of Tx-Rx pairs of the ICGNN plays a crucial role to realize ``plug-and-play" implementation, as the ICGNN does not require re-constructing or re-training for the new Tx-Rx pair.

\section{NUMERICAL RESULTS}

This section provides some numerical results to validate the proposed ICGNN. The structure of the ICGNN under test is given in Table \ref{ICGNN structure}.

\begin{table}[ht]
    \caption{The Architecture of the ICGNN.}
    \label{ICGNN structure}
    \begin{tabular}{c ||c|c|c|c|c|c}
      \hline
      {\#Layer} & {Type} & {IFs} & {OFs} & {AH} & {ReLU} & {BN}\\
      \hline
      \hline
      1 & ${\mathbb C}$GAL & $2 \times{N_{\rm T}}$ & 8 & 8 & {\checkmark} & $\times$\\
      \hline
      2 & ${\mathbb C}$GAL & 64 & 128 & 8 & \checkmark & $\times$ \\
      \hline
      3 & ${\mathbb C}$FL & 1,024 & 1,024 & - & \checkmark & \checkmark\\
      \hline
      4 & ${\mathbb C}$FL & 1,024 & 1,024 & - & \checkmark & \checkmark\\
      \hline
      5 & ${\mathbb C}$FL & 1,024 & 128 & - & \checkmark & \checkmark\\
      \hline
      6 & ${\mathbb C}$FL & 128 & 64 & - & \checkmark & \checkmark\\
      \hline
      7 & ${\mathbb C}$FL & 64 & 1 & - & $\times$ & $\times$\\
      \hline
      8 & ${\mathbb R}$GAL & 1 & 2 & 8 & \checkmark & $\times$ \\
      \hline
      9 & ${\mathbb R}$GAL & 16 & 16 & 8 & \checkmark & $\times$ \\
      \hline
      10 & ${\mathbb R}$GAL & 128 & 128 & 8 & \checkmark & $\times$ \\
      \hline
      11 & ${\mathbb R}$FL & 1,024 & 1,024 & - & \checkmark & \checkmark\\
      \hline
      12 & ${\mathbb R}$FL & 1,024 & 1,024 & - & \checkmark & \checkmark\\
      \hline
      13 & ${\mathbb R}$FL & 1,024 & 128 & - & \checkmark & \checkmark\\
      \hline
      14 & ${\mathbb R}$FL & 128 & 64 & - & \checkmark & \checkmark\\
      \hline
      15 & ${\mathbb R}$FL & 64 & 1 & - & $\times$ & $\times$\\
      \hline
    \end{tabular}
  \begin{tablenotes}
	\footnotesize
	\item IFs: The dimension of the input features.
	\item OFs: The dimension of the output features.
	\item AH: The number of the attention heads.
  \end{tablenotes}
\end{table}

\subsection{Simulation Setting}

\subsubsection{Simulation scenario} The number of Tx-Rx pairs is set as $K\in\{2,3,4,5,6,7,8\}$. The number of antennas of each Tx is set as $N_{\rm T}=4$. The power budget and the constant circuit power are set as $P_{\rm max}=1$ W and $P_{\rm C}=0.1$ W, respectively. The rate requirements of $K$ Rxs are identically set as $1$ bit/s/Hz. The small-scale fading adopts the Rayleigh fading. The average SNR is set as $10$ dB.

Each training sample includes $K\times K$ channel vectors, i.e.,  $\{ {\bf h}_{kj} \}_{k,j\in {\cal K}}$ while each validation or test sample includes $K\times K$ channel vectors and a label which represents the corresponding maximum EE. In this work, we prepare $7$ datasets with different number of Tx-Rx pairs as shown in Table \ref{Datasets}. Particularly, the datasets are categorized into two types.
\begin{itemize}
    \item{Type A:} This type of dataset includes $100,000$ samples which are split into training, validation, and test sets with a ratio of $0.96:0.02:0.02$.
    \item{Type B:} This type of dataset includes $2,000$ test samples only, which is used to test the ICGNN with the unseen problem sizes.
\end{itemize}

\begin{table}[ht]
    \centering
    \caption{Datasets.}
    \label{Datasets}
    \begin{tabular}{c|c|c|c|c}
    \hline
    No. &$N_{\rm T}$ &$K$ & Size& Type\\
     \hline
      \hline
     1   &4&2& 100,000& A\\
     \hline
     2   &4&4& 100,000& A\\
     \hline
     3   &4&6& 100,000& A\\
     \hline
     4   &4&8& 100,000& A\\
     \hline
     5   &4&3& 2,000& B\\
     \hline
     6   &4&5& 2,000& B\\
     \hline
     7   &4&7& 2,000& B\\
     \hline
    \end{tabular}
     \begin{tablenotes}
            \footnotesize
            \item {Type A: The training set, validation set and test set are all included.}
            \item {Type B: Only test set is included.}
    \end{tablenotes}
\end{table}

\subsubsection{Computer configuration} The neural networks are trained and tested under Python 3.8.16 with Pytorch 1.13.1 on a computer with Intel(R) Core(TM) i9-12900K CPU and NVIDIA RTX 3090 (24 GB of memory).

%\subsubsection{Architecture of ICGNN} The detailed architecture of each layer of the proposed ICNet is shown in Table \ref{ICGNN structure}. For the OTA implementation of ICGNN, there are some differences in the architecture of the ${\mathbb R}$GAT part, as shown in Table \ref{OTA structure}.

\subsubsection{Initialization and training} The learnable weights are initialized according to \cite{kaiming_normal} and the learning rate is initialized as $10^{-5}$. The adaptive moment estimation is adopted as the optimizer during the training phase. The batch size is set as $64$ for $500$ epochs of training. The learnable weights with the best performance are used as the training results.

\subsubsection{Performance metrics}  The following metrics are considered to evaluate the ICGNN from the perspectives of model effectiveness and task requirement.
\begin{itemize}
  \item [1)]
  \emph{Optimality performance:} The average ratio of the achievable EE (with feasible solutions) by the DL model to the maximum EE on the test set with identical parameter settings with the training set.
    \item [2)]
 \emph{Scalability performance:} The average ratio of the achievable EE (with feasible solutions) by the DL model to the maximum EE on the test set with different parameter settings from the training set.
  \item [3)]
  \emph{Feasibility rate:} The percentage of the feasible solutions to the considered problem by the DL model.
  \item [4)]
 \emph{Inference time:} The average running time required to calculate the feasible solution with the given channel vectors by the DL model.
\end{itemize}

\subsection{Ablation Experiment}

\begin{table*}[t]
\centering
\caption{Ablation experiment: $4$-antenna $6$-user MISO interference channel system.}
\label{Ablation experiment}
\begin{tabular}{c c c c|c|c || c}
\hline
MP & RD & SR & FE  & OP (Gain) & FR (Gain) & Inference time \\
 \hline
 \hline
 $\times$ & $\times$ & $\times$ & $\times$ & 84.93\% (-) & 89.35\% (-)& 0.0501 ms\\
 \cline{5-7}
$\checkmark$  & $\times$ & $\times$ & $\times$ & 85.75\% (1.82\%) & 95.10\% (5.75\%) & 0.0674 ms\\
 \cline{5-7}
$\checkmark$  & $\checkmark$ & $\times$ & $\times$ & 86.70\% (0.95\%) & 95.85\% (0.75\%) & 0.0684 ms\\
 \cline{5-7}
$\checkmark$ & $\checkmark$ & $\checkmark$ & $\times$ & 93.25\% (6.55\%) & 100\% (4.15\%) & 0.0804 ms\\
 \cline{5-7}
$\checkmark$ & $\checkmark$ & $\checkmark$ & $\checkmark$ & 94.10\% (0.85\%) & 100\% (0.00\%) & 0.0804 ms\\
 \hline
\end{tabular}
\begin{tablenotes}
        \footnotesize
        \item MP/RD/SR/FE: message passing/residual/subgraph representation/feature enhancement.
         \item OP/FR: Optimality performance/feasibility rate.
\end{tablenotes}
\end{table*}

Table \ref{Ablation experiment} presents the results of the ablation experiment to show the message passing, residual connection, graph representation and feature enhancement in the ICGNN in a $(N_{\rm T},K)=(4,6)$ scenario (Dataset No. 3). It is observed that the four mechanisms are all able to improve the optimality performance and feasible rate of the DL model with a limited increase of the inference time. Moreover, in this example, the proposed ICGNN is able to achieve more than $94\%$ optimality with less than $1/10$ millisecond with our computer configuration, which is much more efficient than traditional CVX optimization algorithms.

%As we progressively add the message passing, feature extraction, residual, and graph representation modules, the performance steadily improves, demonstrating that each module positively impacts the model's performance. Considering the inference time, it is evident that the model can achieve near-optimal performance with millisecond-level response times.

\begin{table*}[t]
\belowrulesep=0pt
\aboverulesep=0pt
\centering
\caption{The effectiveness of the hybrid method and model.}
\label{The effectiveness of the hybrid method}
\begin{tabular}{c||c c|c c|c c}
\hline
  & \multicolumn{2}{c|}{End-to-end beamforming} & \multicolumn{2}{c|}{MMSE} & \multicolumn{2}{c}{Hybrid MRT and ZF} \\
\cmidrule(lr){2-3}\cmidrule(lr){4-5}\cmidrule(lr){6-7}
  & OP & FR & OP & FR & OP & FR \\
 \hline
 \hline
 MLP & 54.16\% & 3.45\% & 81.97\% & 50.95\% & 90.66\% & 73.06\%\\
 \hline
 ICGNN & 68.95\% & 42.15\% & 84.48\% & 100\% & 94.10\% & 100\%\\
 \hline
\end{tabular}
\label{table:ablation}
\end{table*}

Table \ref{The effectiveness of the hybrid method} compares the end-to-end beamforming learning\footnote{Via the end-to-end beamforming learning, the DL model is required to directly yield the beamforming vectors, e.g., \cite{icnet}.}, the minimum mean square error (MMSE) learning\footnote{Via the MMSE learning, the DL model is required to yield the power allocation with given the MMSE directions, e.g., \cite{mmselearning}.} and the proposed hybrid MRT and ZF learning. The results of the MLP are also given as a baseline. It is observed that both the MMSE learning and hybrid MRT and ZF learning significantly outperform the end-to-end beamforming learning. The reason is that with prior knowledge, the complexity of the learning task is reduced. Particularly, the output ports of the DL model are reduced from $K\times N_{\rm T}$ complex numbers by the end-to-end beamforming learning to $K$ real numbers by the MMSE learning or $2K$ real numbers by the hybrid MRT and ZF learning. In addition, the hybrid MRT and ZF learning is superior to the MMSE learning due to a more degree of freedom in designing the directions of the beamforming vectors, which expands the capability of the DL model. That is, a trade-off exists between the complexity and capability of the DL model, which is well balanced by the hybrid MRT and ZF learning.  Furthermore, the ICGNN achieves better optimality performance and feasibility rate than the MLP for all schemes due to its more powerful feature extracting modules.

%the effectiveness of the hybrid method compared to directly mapping the beamforming vector or using the MMSE method across different neural network models. From the table, it is evident that the performance of MMSE is significantly higher than that of direct mapping of the beamforming vector. This indicates that mapping only the power of the beamforming vector significantly reduces the model's learning difficulty. The hybrid method further outperforms MMSE because the direction of the beamforming vector is represented as a weighted sum of MRT and ZF directions. The model additionally learns a hybrid coefficient $\alpha$, which raises the performance ceiling.

\subsection{Optimality, Scalability and Feasibility rate}

\begin{table*}[t]
\belowrulesep=0pt
\aboverulesep=0pt
\centering
\caption{Optimality, scalability, and feasibility rate performance evaluation of ICGNN.}
\label{Generalization performance evaluation}
\begin{tabular}{c||c c|c c|c c}
\hline
\multirow{2}{*}{} & \multicolumn{2}{c|}{$K_{\rm {Tr}}=2$} & \multicolumn{2}{c|}{$K_{\rm {Tr}}=4$} & \multicolumn{2}{c}{$K_{\rm {Tr}}=6$} \\

   & SC/OP & FR & SC/OP & FR & SC/OP & FR \\
 \hline
 \hline
 $K_{\rm {Te}}=2$ & {99.92\%}$^\dag$ & {100\%} & 90.50\% & 100\% & 83.28\% & 100\%\\
 $K_{\rm {Te}}=3$ & 94.91\% & 99.70\% & 93.25\% & 99.95\% & 86.99\% & 100\%\\
 $K_{\rm {Te}}=4$ & 80.33\% & 81.80\% & {89.02\%}$^\dag$ & {99.51\%} & 80.02\% & 99.75\%\\
 $K_{\rm {Te}}=5$ & 79.83\% & 82.00\% & 91.61\% & 58.90\% & 90.45\% & 100\%\\
 $K_{\rm {Te}}=6$ & 73.02\% & 75.55\% & 89.66\% & 36.70\% & {94.10\%}$^\dag$ &  100\%\\
 $K_{\rm {Te}}=7$ & 66.48\% & 65.30\% & 86.84\% & 23.95\% & 93.14\% & 98.90\%\\
 $K_{\rm {Te}}=8$ & 61.24\% & 48.75\% & 83.32\% & 12.50\% & 92.42\% & 51.00\%\\
 \hline
\end{tabular}
\begin{tablenotes}
        \footnotesize
        \item $K_{\rm {Tr}}$/$K_{\rm {Te}}$: Value of  $K$ in the training/test set.
        \item OP/SC/FR: Optimality performance/scalability performance/feasibility rate.
        \item{$^\dag$: The result marked with $^\dag$ represents the optimality performance, while the result without $^\dag$ represents the scalability performance.}
\end{tablenotes}
\end{table*}

Table \ref{Generalization performance evaluation} evaluates the optimality, scalability and feasibility rate of the ICGNN. The ICGNN respectively trained with $K=2$ (Dataset No. 1), $K=4$ (Dataset No. 2) and $K=6$ (Dataset No. 3) is tested with $K\in\{2,3,4,5,6,7,8\}$ (Dataset No. 1$\sim$7). That is, the ICGNN is tested in some scenarios with Tx-Rx pairs unseen in the training set. It is observed that the ICGNN is able to achieve good optimality and feasible rate when $K_{\rm Te} = K_{\rm Tr}$. As for the scalability, the ICGNN performs well when $K_{\rm Te}$ is close to $K_{\rm Tr}$. Such a result is appealing, as it allows the operator to directly activate/inactivate several Tx-Rx pairs in the interference channel system without re-training the ICGNN, which greatly enhances the flexibility of resource allocation. However, it is also observed that when $K_{\rm Te}$ is significant different from $K_{\rm Tr}$, the scalability and feasibility rate may be degraded. It is reasonable as the generalization error is inevitable by the DL models. Nevertheless, as mentioned in Remark \ref{transfer}, such an issue can be addressed by using the transfer learning to fine-tune the ICGNN.

%The ICGNN exhibits only minor differences in EE compared to the baseline optimization algorithm across the three training parameter settings, with CSR consistently above 99.5\%. Notably, with $K_{\rm {Tr}}=2$, the performance nearly reaches 100\%. In terms of generalization, the performance remains above 90\% when the number of users is either increased or decreased by one. When the number of users decreases, the CSR does not drop, maintaining a level above 99.5\%.

%描述结果

% 这个效果被证实之后好处（分析下扩展能力）

%unseen problem sizes离training比较远，效果可能就不好

\subsection{Transfer Learning}

\begin{figure}[t]
    \includegraphics[width=0.5\textwidth]{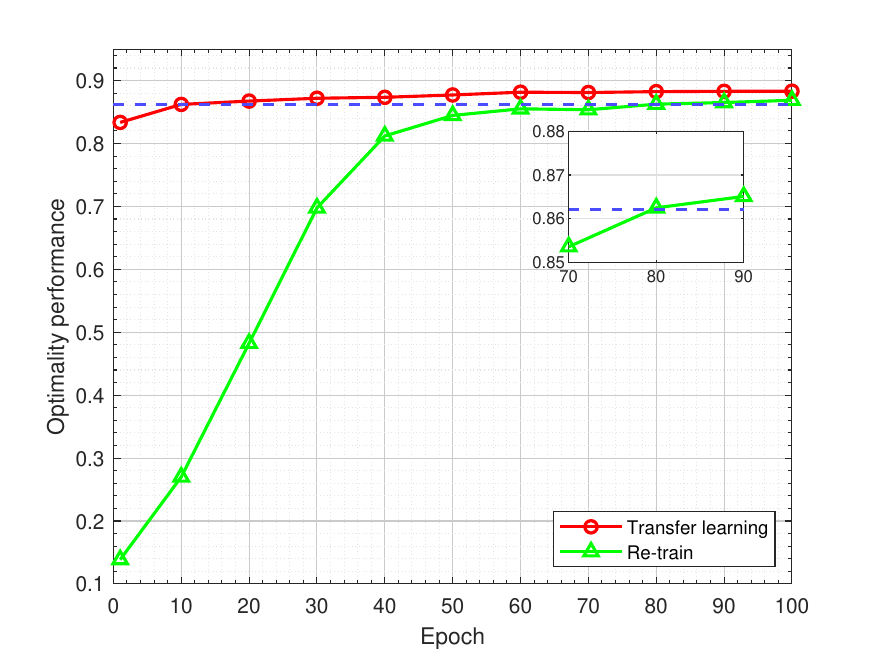}
    \caption{Convergence behavior of the transfer learning ($K_{\rm {Tr}}^{\rm (Pre)}=6$) and re-train.}
    \label{Convergence behavior of the transfer learning and training}
\end{figure}

\begin{table*}[t]
\belowrulesep=0pt
\aboverulesep=0pt
\centering
\caption{Transfer learning: $4$-antenna $8$-user MISO interference channel system.}
\label{Transfer learning}
\begin{threeparttable}
\begin{tabular}{c|c||c c|c c|c c}
\hline
 \multicolumn{2}{c||}{\multirow{2}{*}{}} & \multicolumn{2}{c|}{$K_{\rm {Tr}}^{\rm (Pre)}=2,K_{\rm {Tr}}=8$} & \multicolumn{2}{c|}{$K_{\rm {Tr}}^{\rm (Pre)}=4,K_{\rm {Tr}}=8$} & \multicolumn{2}{c}{$K_{\rm {Tr}}^{\rm (Pre)}=6,K_{\rm {Tr}}=8$} \\
 \cmidrule(lr){3-4}\cmidrule(lr){5-6}\cmidrule(lr){7-8}
 \multicolumn{2}{c||}{} & SC$^\ddag$ & FR & SC & FR & SC & FR \\
 \hline
 \hline
 \multicolumn{2}{c||}{w/o transfer learning} & 61.24\% & 48.75\% & 83.57\% & 12.50\% & 92.36\% & 51.00\% \\
 \hline
  \multirow{4}{*}{\makecell[c]{Transfer\\learning}} & 10 epochs  & 76.97\%  & 100\% & 82.79\% & 99.90\% & 86.21\% & 100\% \\
 \cmidrule(lr){2-2}\cmidrule(lr){3-4}\cmidrule(lr){5-6}\cmidrule(lr){7-8}
  & Gain  & +15.73\% & +51.25\% & -0.78\% & +87.50\% & -6.15\% & +49.00\% \\
 \cline{2-8}
  & 500 epochs  & 89.11\% & 100\% & 89.85\% & 100\% & 90.12\% & 100\% \\
 \cmidrule(lr){2-2}\cmidrule(lr){3-4}\cmidrule(lr){5-6}\cmidrule(lr){7-8}
  & Gain  & +27.87\% & +51.25\% & +6.28\% & +87.50\% & -2.24\% & +49.00\% \\
\hline
 \multicolumn{2}{c||}{\multirow{3}{*}{Re-train}} & \multicolumn{6}{c}{$K_{\rm {Tr}}=8$}  \\
 \cmidrule(lr){3-8}
 \multicolumn{2}{c||}{} & \multicolumn{3}{c}{OP} & \multicolumn{3}{c}{FR} \\
 \cline{3-8}
 \multicolumn{2}{c||}{} & \multicolumn{3}{c}{89.08\%} & \multicolumn{3}{c}{100\%} \\
\hline
\end{tabular}
\begin{tablenotes}
    \footnotesize
    \item $K_{\rm {Tr}}^{\rm (Pre)}$: Value of $K$ during pre-training.
     \item{$^\ddag$: For transfer learning, scalability performance stands for the average ratio of the achievable EE (with feasible solutions) by the DL model to the maximum EE on the test set with different parameter settings from the pre-training set.}
\end{tablenotes}
\end{threeparttable}
\end{table*}

%retrain和direct learning

Fig. \ref{Convergence behavior of the transfer learning and training} show the convergence behavior of the ICGNN via transfer learning and re-training with $K=8$ (Dataset No. 4). For the transfer learning, the learnable parameters are initialized by that of the well-trained ICGNN with $K=6$ (Dataset No. 3), while for the re-training the learnable parameters are initialized according to \cite{kaiming_normal}. It is observed that the transfer learning achieves a much faster convergence speed than that of the re-training. Particular, the transfer learning after $10$ epochs achieves the commensurate performance with re-training after $80$ epochs. Therefore, the transfer learning can save the training time.

%the performance of transfer learning after 10 epochs approaches that of direct learning after approximately 80 epochs. This indicates that transfer learning not only offers faster training speeds and higher performance than direct training but also provides a solution for scenarios with poor generalization performance.

Table \ref{Transfer learning} shows the scalability performance and feasibility rate of the ICGNN via transfer learning with $K=8$ (Dataset No. 4). The learnable parameters are initialized by the corresponding values for the well-trained ICGNN with $K=2$ (Dataset No. 1), $K=4$ (Dataset No. 2) and $K=6$ (Dataset No. 3), respectively. For comparison, the results of the ICGNN via re-training with $K=8$ (Dataset No. 4) and the generalization results of the ICGNN via re-training with $K=2$ (Dataset No. 1), $K=4$ (Dataset No. 2) and $K=6$ (Dataset No. 3) are also given. It is observed that the results of the transfer learning with $10$ epochs are already competitive. Especially, the feasibility rate is greatly enhanced. Combining with Fig. \ref{Convergence behavior of the transfer learning and training}, the transfer learning is able to achieve notable performance gain with limited fine-tune cost. Moreover, the transfer learning with $500$ epochs is able to achieve better performance than re-training.

\subsection{Over-the-air  Distributed Implementation}

\begin{figure}[t]
    \includegraphics[width=0.5\textwidth]{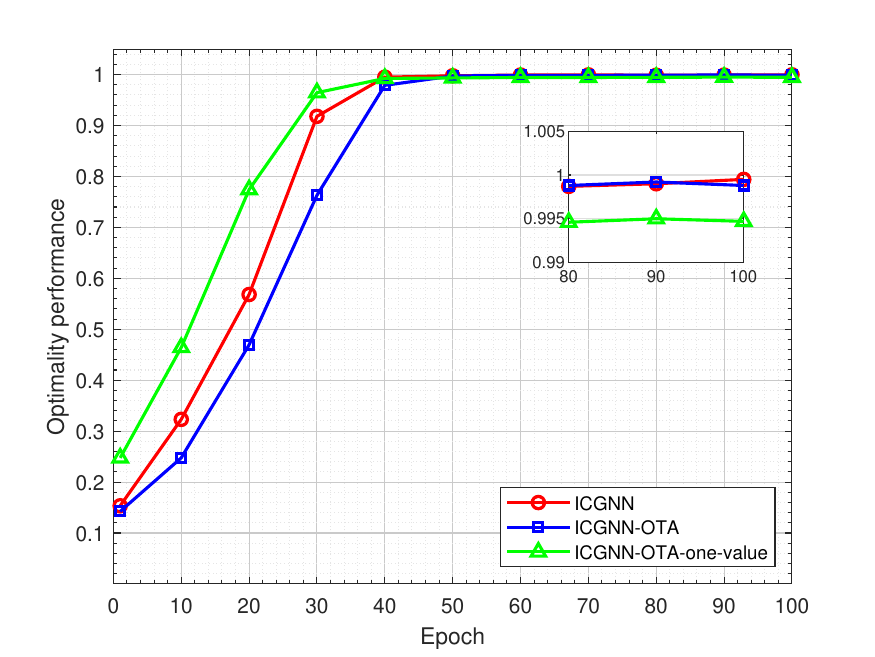}
    \caption{The convergence behavior and performance of ICGNN after implementing over-the-air.}
    \label{over-the-air performance 2}
\end{figure}

\begin{figure}[t]
    \includegraphics[width=0.5\textwidth]{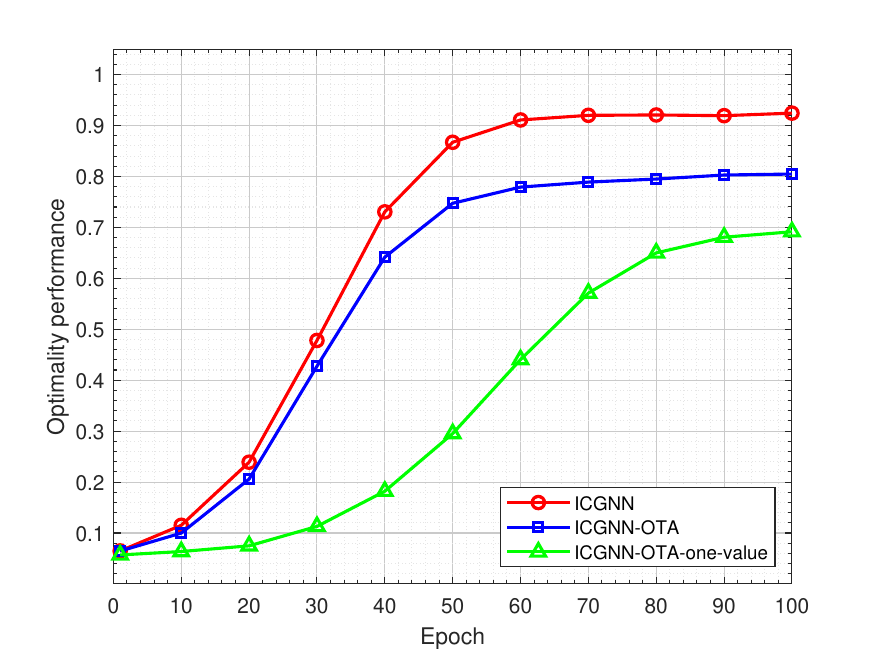}
    \caption{The convergence behavior and performance of ICGNN after implementing over-the-air.}
    \label{over-the-air performance 6}
\end{figure}

Fig. \ref{over-the-air performance 2} and Fig. \ref{over-the-air performance 6} illustrate the convergence behaviors of the ICGNN via centralized learning, OTA distributed learning and one-value OTA distributed learning, respectively. Particularly, for the centralized learning, all Tx-Rx pairs share the learnable parameters. for the OTA distributed learning, each Tx-Rx pair maintains its local learnable parameters; for the one-value OTA distributed learning, each Tx-Rx pair not only maintains its local learnable parameters but also broadcasts one real number during the message passing. In Fig. \ref{over-the-air performance 2} and Fig. \ref{over-the-air performance 6}, $K$ is set as $2$ (Dataset No. 1) and $6$ (Dataset No. 3), respectively. It is observed that the performances of the centralized learning and distributed learning are very close for $K=2$ while the centralized learning outperforms the distributed learning for $K=6$. The reason is that the total number of the learnable parameters increases with $K$ via the distributed learning as there is no parameter sharing. Nevertheless, the distributed learning enables ``plug-and-play" implementation which makes the interference channel system more scalable. Furthermore, in Fig. \ref{over-the-air performance 6}, the ICGNN via  OTA distributed learning is superior to that via one-value OTA distributed learning at the cost of more signaling overhead. That is, a trade-off is revealed between the learning performance and the signaling overhead.

%缺少两种distributed learning的实现方式

%and performance of ICGNN before and after implementing OTA deployment in scenarios with 2 users and 6 users respectively.illustrates the convergence behavior and performance of ICGNN before and after implementing OTA deployment in a scenario with 6 users and 8 antennas. In the 2-user scenario, the performance after OTA deployment is essentially consistent with that before deployment. However, in the 6-user scenario, there is a performance decline. This performance degradation can be attributed to the lack of weight-sharing capability between different computational nodes in the OTA implementation, resulting in a lower level of optimization in training results compared to before OTA implementation. If the amount of transmitted data is further reduced to save signaling overhead, the performance will decrease further due to the reduction in the number of neurons.

\section{Conclusion}
This work proposed a scalable beamforming scheme for MISO interference channels, termed ICGNN, which enhanced the learning performance via several novel designs, including hybrid MRT and ZF scheme, feature enhancement module, SGR, multi-head attention, and residual connection. With OTA, the ICGNN can be realized via distributed implementation and support the ``plug-and-play" implementation of a new Tx-Rx pair to an existing wireless network. Experimental results demonstrated that the proposed ICGNN achieved near-optimal performance with an inference time less than $0.1$ ms. Moreover, ICGNN was capable of handling varying numbers of Tx-Rx pairs, and its generalization capability was further enhanced through transfer learning. Moreover, the trade-off between the learning performance and the signaling overhead was revealed for the distributed implementation.


\begin{thebibliography}{10}

\bibitem{6G}
C. -X. Wang et al., ``On the road to 6G: visions, requirements, key technologies, and testbeds," \emph{IEEE Commun. Surv. Tutorials}, vol. 25, no. 2, pp. 905-974, Second quarter 2023.

\bibitem{intro2}
Y. Lu, K. Xiong, P. Fan, Z. Ding, Z. Zhong, and K. B. Letaief, "Global energy efficiency in secure MISO SWIPT systems with non-linear power-splitting EH model," \emph{IEEE J. Sel. Areas Commun.}, vol. 37, no. 1, pp. 216-232, Jan. 2019.

\bibitem{intro3}
Y. Lu, K. Xiong, P. Fan, B. Ai, and Z. Zhong, ``Outage-constrained sum transmission rate maximization in RIS-assisted MISO systems," \emph{IEEE Trans. Wireless Commun.}, vol. 23, no. 4, pp. 2505-2518, Apr. 2024.

\bibitem{intro4}
Y. Shi et al., ``Machine learning for large-scale optimization in 6G wireless networks," \emph{IEEE Commun. Surv. Tutorials}, vol. 25, no. 4, pp. 2088-2132, Fourth quarter 2023.

\bibitem{intro5}
Y. Lu, W. Mao, H. Du, O. A. Dobre, D. Niyato, and Z. Ding, ``Semantic-aware vision-assisted integrated sensing and communication: Architecture and resource allocation," \emph{IEEE Wireless Commun.}, vol. 31, no. 3, pp. 302-308, June 2024.

\bibitem{MLP for beamforming design}
S. Biswas, U. Singh, and K. Nag, ``Multi-layer perceptron-based beamformer design for next-generation full-duplex cellular systems," in Proc. \emph{IEEE/ACIS SNPD}, pp. 49-55, 2021.

\bibitem{CNN for beamforming design}
H. Sun, X. Feng, J. Wang, M. Zhou, and X. Kuai, ``Beamforming design via deep learning for underwater acoustic communications," in Proc. \emph{IEEE ICCT}, pp. 576-580, 2021.

\bibitem{Graph structure}
Y. Shen, J. Zhang, S. H. Song, and K. B. Letaief, ``Graph neural networks for wireless communications: From theory to practice," \emph{IEEE Trans. Wireless Commun.}, vol. 22, no. 5, pp. 3554-3569, May 2023.

\bibitem{Graph neural networks survey}
H. Jiang, X. Xing, S. Li, J. Chen, and Z. Jia, ``Graph neural networks based recommendation methods in different scenarios: A survey," in Proc. \emph{ICNISC}, pp. 660-666, 2022.

\bibitem{lugnn}
Y. Lu et al., ``Graph neural networks for wireless networks: Graph representation, architecture and evaluation," arXiv:2404.11858.

\bibitem{rw1}
V. -C. Luu and J. -P. Hong, ``GNN-based meta-Learning approach for adaptive power control in dynamic D2D communications," \emph{IEEE Trans. Veh. Technol.}, vol. 73, no. 4, pp. 5982-5987, April. 2024.

\bibitem{rw3}
M. Eisen and A. Ribeiro, ``Large scale wireless power allocation with graph neural networks," \emph{IEEE International Workshop SPAWC}, pp. 1-5, 2019.

\bibitem{rw2}
M. Eisen and A. Ribeiro, ``Optimal wireless resource allocation with random edge graph neural networks," \emph{IEEE Trans. Signal Process.}, vol. 68, pp. 2977-2991, 2020.

\bibitem{cw1}
T. Chen, M. You, G. Zheng, and S. Lambotharan, ``Graph neural network based beamforming in D2D wireless networks," in Proc. \emph{WSA}, pp. 1-5, 2021.

\bibitem{cw4}
Y. Shen, Y. Shi, J. Zhang, and K. B. Letaief, ``Graph neural networks for scalable radio resource management: Architecture design and theoretical analysis," \emph{IEEE J. Sel. Areas Commun.}, vol. 39, no. 1, pp. 101-115, Jan. 2021.

\bibitem{cw5}
T. Jiang, H. V. Cheng, and W. Yu, ``Learning to reflect and to beamform for intelligent reflecting surface with implicit channel estimation," \emph{IEEE J. Sel. Areas Commun.}, vol. 39, no. 7, pp. 1931-1945, July 2021.

\bibitem{dl}
Y. Li, Y. Lu, R. Zhang, B. Ai, and Z. Zhong, ``Deep learning for energy efficient beamforming in MU-MISO networks: A GAT-based approach," \emph{IEEE Wireless Commun. Lett.}, vol. 12, no. 7, pp. 1264-1268, July 2023.

\bibitem{dl3}
Y. Li, Y. Lu, B. Ai, Z. Zhong, D. Niyato, and Z. Ding, ``GNN-enabled max-min fair beamforming," \emph{IEEE Trans. Veh. Technol.}, vol. 73, no. 8, pp. 12184-12188, Aug. 2024.

\bibitem{dl2}
Y. Li, Y. Lu, B. Ai, O. A. Dobre, Z. Ding, and D. Niyato, ``GNN-based beamforming for sum-rate maximization in MU-MISO networks," \emph{IEEE Trans. Wireless Commun.},  vol. 23, no. 8, pp. 9251-9264, Aug. 2024.

\bibitem{sc2}
L. Yu, J. Yuan, X. Lv, Z. Wang, D. Cai, and Z. Dong, ``Power allocation optimization design for D2D cellular networks based on graph neural networks," in Proc. \emph{ICCT}, pp. 564-569, 2023.

\bibitem{sc3}
J. Kim, H. Lee, S. -E. Hong, and S. -H. Park, ``A bipartite graph neural network approach for scalable beamforming optimization," \emph{IEEE Trans. Wireless Commun.}, vol. 22, no. 1, pp. 333-347, Jan. 2023.

\bibitem{sc4}
Y. Wang, Y. Li, Q. Shi, and Y. -C. Wu, ``Learning cooperative beamforming with edge-update empowered graph neural networks," in Proc. \emph{ICC}, pp. 5111-5116, 2023.

\bibitem{dt1}
Z. Wang, M. Eisen, and A. Ribeiro, ``Decentralized wireless resource allocation with graph neural networks," in Proc. \emph{IEEE ACSSC}, pp. 299-303, 2020.

\bibitem{dt2}
Z. Zhao, G. Verma, C. Rao, A. Swami, and S. Segarra, ``Distributed scheduling using graph neural networks," in Proc. \emph{ICASSP}, pp. 4720-4724, 2021.

\bibitem{dt3}
Z. Zhao, A. Swami, and S. Segarra, ``Distributed link sparsification for scalable scheduling using graph neural networks," in Proc. \emph{ICASSP}, pp. 5308-5312, 2022.

\bibitem{ota}
Y. Gu, C. She, Z. Quan, C. Qiu, and X. Xu, ``Graph neural networks for distributed power allocation in wireless networks: Aggregation over-the-air," \emph{IEEE Trans. Wireless Commun.}, vol. 22, no. 11, pp. 7551-7564, Nov. 2023.

\bibitem{hyb}
 W. Guo, Y. Lu, H. Du, B. Ai, D. Niyato, and Z. Ding, ``Hybrid MRT and ZF learning for energy-efficient transmission in multi-RIS-assisted networks," \emph{IEEE Trans. Veh. Technol.}, vol. 73, no. 8, pp. 12247-12251, Aug. 2024.

\bibitem{Interference Channel}
W. -C. Li, T. -H. Chang, C. Lin and C. -Y. Chi, ``Coordinated beamforming for multiuser MISO interference channel under rate outage constraints," \emph{IEEE Trans. Signal Process.}, vol. 61, no. 5, pp. 1087-1103, Mar. 2013.

 \bibitem{Jorswieck}
E. A. Jorswieck, E. G. Larsson, and D. Danev, ``Complete characterization of the pareto boundary for the MISO interference channel," \emph{IEEE Trans. Signal Process.}, vol. 56, no. 10, pp. 5292-5296, Oct. 2008.

\bibitem{heterogeneit}
X. Wang et al., ``Heterogeneous graph attention network,” in Proc. \emph{ACM WWW}, pp. 2022-2032, 2019.

\bibitem{BatchNorm}
S. Ioffe and S. Christian, ``Batch normalization: Accelerating deep network training by reducing internal covariate shift," in Proc. \emph{ICML}, 2015.

\bibitem{Adam}
D. P. Kingma and J. Ba, ``Adam: A method for stochastic optimization,” in Proc. \emph{ICLR}, pp. 1–15, Feb. 2015.

\bibitem{kaiming_normal}
K. He, X. Zhang, S. Ren, and J. Sun, ``Delving deep into rectifiers: Surpassing human-level performance on ImageNet classification," in Proc. \emph{ICCV}, pp. 1026-1034, 2015.

\bibitem{icnet}
C. He, Y. Li, Y. Lu, B. Ai, Z. Ding, and D. Niyato, ``ICNet: GNN-enabled beamforming for MISO interference channels with statistical CSI," \emph{IEEE Trans. Veh. Technol.}, vol. 73, no. 8, pp. 12225-12230, Aug. 2024.

\bibitem{mmselearning}
C. Hu, Y. Lu, H. Du, M. Yang, B. Ai, and D. Niyato, ``AI-empowered RIS-assisted networks: CV-enabled RIS selection and DNN-enabled transmission," early accessed \emph{IEEE Trans. Veh. Technol.}, 2024.


% \bibitem{phsec}
% Y. Zhang, Y. Lu, R. Zhang, B. Ai, and D. Niyato, ``Deep reinforcement learning for secrecy energy efficiency maximization in RIS-assisted networks," \emph{IEEE Trans. Veh. Technol.}, vol. 72, no. 9, pp. 12413-12418, Sept. 2023.

% \bibitem{cv-channel}
% Z. Hua, Y. Lu, G. Pan, K. Gao, D. B. d. Costa, and S. Chen, ``Computer vision-aided mmWave UAV communication systems," \emph{IEEE Internet Things J.}, vol. 10, no. 14, pp. 12548-12561, July 2023.


\end{thebibliography}
\end{document}